\documentclass[10pt,journal,compsoc]{IEEEtran}
\ifCLASSOPTIONcompsoc
  \usepackage[nocompress]{cite}
\else
  \usepackage{cite}
\fi
\ifCLASSINFOpdf
  \usepackage[pdftex]{graphicx}
\else
  \usepackage[dvips]{graphicx}
\fi
\usepackage{amsmath}
\usepackage{amsfonts}
\usepackage{amssymb}
\usepackage{mathtools}
\usepackage{algorithm}
\usepackage{algpseudocode}

\usepackage{wrapfig}
\usepackage{multirow}
\usepackage{makecell}
\usepackage{enumitem}
\usepackage[hidelinks]{hyperref}
\usepackage{microtype}

\DeclareMathOperator{\Tr}{tr}

\begin{document}
%
\title{Parallel and Scalable Heat Methods for Geodesic Distance Computation}
%
%
%
%

\author{
	Jiong~Tao,
	Juyong~Zhang$^\dagger$,~\IEEEmembership{Member,~IEEE,}
	Bailin~Deng,~\IEEEmembership{Member,~IEEE,}\\
	Zheng~Fang,
	Yue~Peng,
	and Ying~He,~\IEEEmembership{Member,~IEEE}
	\IEEEcompsocitemizethanks{\IEEEcompsocthanksitem J. Tao, J. Zhang and Y. Peng are with School of Mathematical Sciences, University of Science and Technology of China.
	\IEEEcompsocthanksitem B. Deng is with  School of Computer Science and Informatics, Cardiff University.
	\IEEEcompsocthanksitem Z. Fang and Y. He are with School of Computer Science and Engineering, Nanyang Technological University.}
	\thanks{$^\dagger$Corresponding author. Email: \texttt{juyong@ustc.edu.cn}.}
}

%
%

\markboth{~}
{~}
%



\IEEEtitleabstractindextext{%
\begin{abstract}
	In this paper, we propose a parallel and scalable approach for geodesic distance computation on triangle meshes. Our key observation is that the recovery of geodesic distance with the heat method~\cite{CraneWW13} can be reformulated as optimization of its gradients subject to integrability, which can be solved using an efficient first-order method that requires no linear system solving and converges quickly. Afterward, the geodesic distance is efficiently recovered by parallel integration of the optimized gradients in breadth-first order. Moreover, we employ a similar breadth-first strategy to derive a parallel Gauss-Seidel solver for the diffusion step in the heat method. To further lower the memory consumption from gradient optimization on faces, we also propose a formulation that optimizes the projected gradients on edges, which reduces the memory footprint by about 50\%. Our approach is trivially parallelizable, with a low memory footprint that grows linearly with respect to the model size. This makes it particularly suitable for handling large models. Experimental results show that it can efficiently compute geodesic distance on meshes with more than 200 million vertices on a desktop PC with 128GB RAM, outperforming the original heat method and other state-of-the-art geodesic distance solvers.
\end{abstract}

\begin{IEEEkeywords}
Heat method, heat diffusion, Poisson equation, scalability, parallel algorithm.
\end{IEEEkeywords}}

\maketitle

\IEEEdisplaynontitleabstractindextext

%
\IEEEpeerreviewmaketitle

\IEEEraisesectionheading{\section{Introduction}}
\label{sec:intro}

\IEEEPARstart{G}{eodesic} distance is a commonly used feature of geometry and has a wide range of applications in computer vision and computer graphics~\cite{PeyrePKC10}. For example, geodesic distance provides an expression-invariant representation of human faces that can be used for 3D face recognition~\cite{BronsteinBK05}. Other important applications include object segmentation and tracking~\cite{Najman:1996,ParagiosD00,WangSYP18}, shape analysis~\cite{BrynerKLS14}, and texture mapping~\cite{ZigelmanKK02}.

Many algorithms have been proposed to compute geodesic on polyhedral meshes, like fast matching and fast sweeping~\cite{Sethian96,Sethian99,Kimmel1998}, the Mitchell-Mount-Papadimitriou (MMP) algorithm~\cite{Mitchell87}, and the Chen-Han (CH) algorithm~\cite{Chen90}. Recently, the heat method (HM)~\cite{CraneWW13,Crane2017} was proposed to compute geodesic distances on discrete domains, such as regular grids, point clouds, triangle meshes, and tetrahedral meshes. It is based on Varadhan's formula~\cite{var} that relates the heat kernel and geodesic distance:
\begin{equation}
\lim_{t\rightarrow 0} -4t\log h(t,x,y)=d^2(x,y),
\label{eq:Varadhan}
\end{equation}
where $x,y$ is an arbitrary pair of points on a Riemannian manifold $M$, $t$ is the diffusion time, $h(t,x,y)$ and $d(x,y)$ are the heat kernel and geodesic distance respectively.
The heat method is conceptually simple and elegant.
Since the geodesic distance is a solution to the Eikonal equation
\begin{equation}
\|\nabla d\| = 1,
\label{eq:eikonal}
\end{equation}
the heat method first integrates the heat flow $\dot{u}=\bigtriangleup u$ for a short time and normalizes its gradient to derive a unit vector field that approximates the gradient of the geodesic distance. Afterward, it determines the geodesic distance by finding scalar field $d$ whose gradient is the closest to the unit vector field, which amounts to merely solving a Poisson system.
The heat method involves only sparse linear systems, which can be pre-factorized once and reused to solve different right-hand-sides in linear time.
This feature makes it highly attractive for applications where geodesic distances are required repeatedly.
However, the Cholesky factorization used in~\cite{CraneWW13,Crane2017} requires a substantial amount of computational time and memory for large meshes.
As a result, the heat method does not scale well. Although we can replace the Cholesky direct solver with iterative solvers with lower memory consumption such as Krylov subspace methods, these solvers can still take a large number of iterations with long computational time.

\begin{figure*}[htbp]
	\centering
	\includegraphics[width=\textwidth]{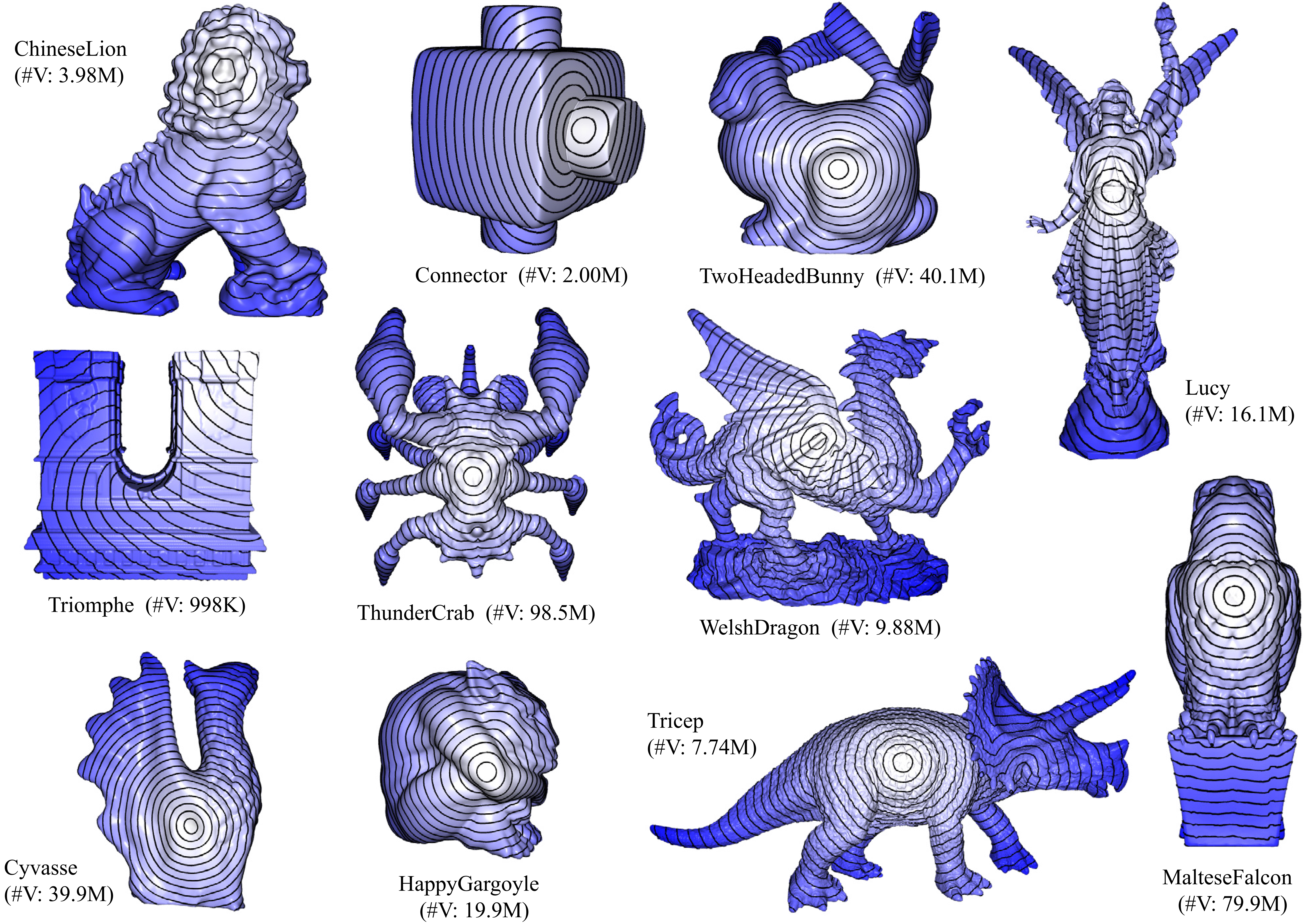}\\
	\caption{Geodesic distance fields computed using our parallel and scalable heat method (Algorithm~\ref{alg:Geodesic}) on different high-resolution models, visualized using their level sets. A comparison with other methods on computational time, peak memory consumption and accuracy is provided in Tab.~\ref{tab:results}.}
	\label{fig:gallery}
\end{figure*}

As technological advances make computational and imaging devices more and more powerful, we can now capture and reconstruct 3D models in higher and higher resolution. Therefore, there will be a growing demand for algorithms that can efficiently handle such large-scale data.
The goal of this work is to develop a scalable algorithm for computing geodesic distance on mesh surfaces. Our method follows a similar approach as the heat method, first computing approximate gradients via heat diffusion and then using them to recover the distance. Our main insight is that instead of solving the Poisson system, we can perform the second step indirectly with much better efficiency and scalability. Specifically, we first compute an integrable gradient field that is closest to the unit vector field derived from heat diffusion, then integrate this field to obtain the geodesic distance. We formulate the computation of gradient field as a convex optimization problem, which can be solved efficiently using the alternating direction method of multipliers (ADMM)~\cite{Boyd2011-admm}, a first-order method with fast convergence. Unlike previous ADMM solvers that optimize function variables to regularize the gradients, our formulation uses the gradients as variables, such that each step of the solver only involves a separable subproblem and is trivially parallelizable. The resulting gradient field can be efficiently integrated using a parallel algorithm. For the heat diffusion step, we also propose a parallel Gauss-Seidel solver, which is more efficient and robust for large meshes than direct and iterative linear solvers. The resulting method is both efficient and scalable, and can be run in parallel to gain speedup on multi-core CPUs and GPUs. We evaluate the performance of our method using a variety of mesh models in different sizes. Our method significantly outperforms the original heat method while achieving similar accuracy (see Fig.~\ref{fig:gallery} and Tab.~\ref{tab:results}). Moreover, the computational time and memory consumption of our approach grow linearly with the mesh size, allowing it to handle much larger meshes than the heat method.

Even though our optimization approach for computing integrable gradients already provides a significant boost to the computational and memory performance compared to the original heat method, its memory footprint can be further reduced. Our key observation is that the closeness between the geodesic distance gradients and their target values can be reformulated as the closeness between their projections on the mesh edges, allowing us to formulate the problem with many fewer variables. To this end, we propose an edge-based formulation that optimizes the change of geodesic distance along the mesh edges, which we solve using a similar ADMM solver. Such changes along edges encode the directional derivatives of the geodesic distance, and the optimization result can be directly used to recover the geodesic distance by integration. Compared to the approach of computing integrable face-based gradients, our edge-based solver can further reduce the memory footprint by about 50\%. This allows us to process a model with over 200 million vertices on a desktop PC with 128GB RAM, where even the face-based optimization approach fails due to excessive memory consumption (see Fig.~\ref{fig:edgebased}).   

\subsection{Our Contributions}
Our main contribution is new approaches that can compute geodesic distance on mesh surfaces in an efficient and scalable way, including:
\begin{itemize}
	\item A parallel Gauss-Seidel solver for solving short-time heat diffusion from source vertices, which is more scalable and numerically more robust than directly solving the heat diffusion linear system in~\cite{CraneWW13}.
	\item A convex optimization formulation for correcting the unit vector field derived from heat diffusion into an integrable gradient field, and an efficient ADMM solver for the problem. Our solver is trivially parallelizable, and has linear time and space complexity.
	\item A novel edge-based formulation that optimizes the change of geodesic distance along mesh edges, which encodes the differential information of geodesic distance in a compact way and further reduces the memory footprint compared to the face-based gradient optimization.   
	\item An efficient parallel integration scheme to recover geodesic distance from either the face-based gradients or the edge-based changes.
\end{itemize}

\section{Related Work}
\label{sec:related-work}
In the past three decades, various techniques have been proposed to compute geodesic distance on mesh surfaces.
A classical approach is to maintain wavefront on mesh edges and propagate it across the faces in a Dijkstra-like sweep. Seminal works include the MMP algorithm~\cite{Mitchell87} and the CH algorithm~\cite{Chen90}, with worst-case running time of $O(n^2 \log n)$ and $O(n^2)$ respectively on a mesh with $n$ vertices. Although these two methods work for arbitrary manifold triangle meshes and can compute exact geodesic distances, they do not scale well to large models due to their high computational complexities. Subsequently, various acceleration techniques have been proposed to achieve more practical performance. Surazhsky et al.~\cite{Surazhsky05} proposed an efficient implementation of the MMP algorithm, and a fast algorithm for geodesic distance approximation. Xin and Wang~\cite{Xin09} improved the performance of the CH algorithm by filtering wavefront windows and maintaining a priority queue. Ying et al.~\cite{Ying13parallel} proposed a parallel CH algorithm that propagates a large number of wavefront windows at the same time. Similarly, Xu et al.~\cite{XuLXJ11} accelerated the wavefront propagation for both the MMP algorithm and the CH algorithm by simultaneous propagation of multiple windows. Qin et al.~\cite{QH2016} proposed a triangle-oriented wavefront propagation algorithm with a window pruning strategy to improve computational speed.

Another class of algorithms, called graph-based methods, first pre-computes a sparse graph that encodes the geodesic information of the surface. Geodesic distance query is then performed by computing a shortest path on the graph. Ying et al.~\cite{SVG} proposed the saddle vertex graph (SVG) for encoding geodesic distance. 
Constructing the SVG takes $O(nK^2\log K)$ time, whereas computing the geodesic distance takes $O(Kn\log n)$ time, where $K$ is the maximal degree. 
Wang et al.~\cite{DGG17} proposed the discrete geodesic graph (DGG) to approximate geodesic distance with given accuracy and empirically linear computational time. Xin et al.~\cite{Xin2018-Lightweight} pre-computed a proximity graph between a set of sample points on the surface, and augmented the graph with the source and target vertices to compute the shortest path and approximate the geodesic distance. For graph-based methods, although the graph construction can potentially be done in parallel, the distance query is intrinsically sequential.

As the geodesic distance function $u$ satisfies the Eikonal equation, various methods have been proposed to solve this PDE for the geodesic distance. The fast marching method (FMM)~\cite{Kimmel1998} solves the equation by iteratively building the solution outward from the points with known/smallest distance values on regular grids and triangulated surfaces, with a running time of $O(n \log n)$ on a triangle mesh with $n$ vertices. Weber et al.~\cite{Weber:2008:PAA:1409625.1409626} proposed an efficient parallel FMM on geometry images, which runs in $O(n)$ time. The heat method proposed by Crane et al.~\cite{CraneWW13} adopts a different strategy: rather than solving the distance function directly, it first computes a unit vector field that approximates its gradient, then integrates this vector field by solving a Poisson equation; this only requires solving two linear systems and is highly efficient. This approach was recently generalized to compute parallel transport vector-valued data on a curved manifold~\cite{Sharp2019}. 
Our approach is also based on the heat method, but using iterative solvers for the linear systems instead of the direct solvers in~\cite{CraneWW13}. The adoption of iterative solvers not only reduces the memory footprint and makes it easier to parallelize the algorithm, but also allows the user to stop the computation early in cases where lower accuracy but higher performance is needed; such early termination is not possible with direct solvers, as they always carry through the computation exactly.
The heat method was also extended by Belyaev and Fayolle~\cite{belyaev2015} to compute the general $L_p$ distance to a 2D curve or a 3D surface via a non-convex constrained optimization, using ADMM as the numerical solver. Although our method also employs ADMM to solve an optimization problem, it is fundamentally different from the approach in~\cite{belyaev2015}. Unlike~\cite{belyaev2015}, we solve a convex optimization problem, for which there is a strong convergence guarantee with ADMM~\cite{Boyd2011-admm}. Moreover, as the formulation in~\cite{belyaev2015} optimizes the distance function values, their ADMM solver still involves linear system solving in each iteration, thereby suffering from the same limitation in scalability as the original heat method. Our approach optimizes the gradient of the geodesic distance instead, which allows our ADMM solver to avoid linear system solving and achieve better efficiency and scalability.   
\section{Algorithm}
\label{sec:algorithm}

In this paper, we consider geodesic distance computation on a triangle mesh $\mathcal{M} = (\mathcal{V}, \mathcal{E}, \mathcal{F})$, where $\mathcal{V}$, $\mathcal{E}$, and $\mathcal{F}$ denote the set of vertices, edges, and faces, respectively.
Given a source vertex $v_s$, we want to compute the geodesic distance $d_i$ from each vertex $v_i$ to $v_s$ by solving the Eikonal equation with boundary condition $d(v_s) = 0$.
Inspired by the heat method~\cite{CraneWW13}, we develop a gradient-based solver for the Eikonal equation.
Similar to the heat method, we first construct a unit vector $\{\mathbf{h}_i\}$ for each face $f_i$ using short-time heat diffusion from the source vertex. Such a vector field provides a good approximation to the gradient field of the geodesic distance function, but is in general not integrable. The geodesic distance is then computed as a scalar field whose gradient is as close to $\mathbf{h}_i$ as possible. Different from the original heat method that computes heat diffusion and recovers geodesic distance by directly solving linear systems, we perform these steps using iterative methods: heat diffusion is done via a Gauss-Seidel iteration (Section~\ref{sec:heat}), whereas the geodesic distance is computed by first correcting $\{\mathbf{h}_i\}$ into an integrable field $\{\mathbf{g}_i\}$ with an ADMM solver (Section~\ref{sec:integrable}) and then integrating it from the source vertex (Section~\ref{sec:integration}). Algorithm~\ref{alg:Geodesic} shows the pipeline of our method. The main benefit of our approach is its scalability: both the Gauss-Seidel heat diffusion and the geodesic distance integration can be performed in parallel through breadth-first traversal over the mesh surface; the correction of $\{\mathbf{h}_i\}$ is a simple convex optimization problem and computed using an ADMM solver where each step is trivially parallelizable. Moreover, both the Gauss-Seidel solver and the ADMM solver converge quickly to a solution with reasonable accuracy, resulting in much less computational time than the original heat method. Finally, our approach has a low memory footprint that grows linearly with mesh size and allows it to handle very large meshes, while the original heat method fails on such models due to the memory consumption of matrix factorization.

\begin{algorithm}[t]
	\caption{A parallel and scalable heat method via face-based gradient optimization}\label{alg:Geodesic}
	\begin{algorithmic}[1]
		\Require $\mathcal{M} = (\mathcal{V}, \mathcal{E}, \mathcal{F})$: a manifold triangle mesh;~
		$v_s$: the source vertex;~
		$t$: heat diffusion time.
		\State $\{u_i | {i\in \mathcal{V}}\} =  \mathrm{Diffusion}(M, v_s, t)$ \Comment{\textsf{Heat diffusion from source vertex}}
		\State $\mathbf{H} = - {\nabla u}/{\|\nabla u\|}$ \Comment{\textsf{Face-wise gradient normalization}}
		\State $\mathbf{G} =  \textrm{Integrable}(M, \mathbf{H})$ \Comment{\textsf{Optimize an integrable gradient field}}
		\State $\{d_i | i\in \mathcal{V}\} =  \textrm{Recovery}(M, v_s, \mathbf{G})$ \Comment{\textsf{Recover geodesic distance}}
	\end{algorithmic}
\end{algorithm}

To facilitate presentation, we assume for now that the triangle mesh $\mathcal{M}$ is a topological disk, with only one source vertex. More general cases, such as meshes of arbitrary topology and multiple sources, will be discussed in Section~\ref{sec:extensions}.

\subsection{Heat Diffusion}
\label{sec:heat}
It is shown in~\cite{CraneWW13} that approximating the geodesic distance using the Varadhan's formula~\eqref{eq:Varadhan} directly will yield poor results due to its high sensitivity to errors in magnitude. Instead, the heat method exploits its connection with heat diffusion to approximate the gradient of geodesic distance.  
Following this approach, we compute the initial vector field $\{\mathbf{h}_i\}$ by integrating the heat flow $\dot{u} = \Delta u$ of a scalar field $u$ for a short time $t$, and taking
\begin{equation}
	\mathbf{h}_i = - \frac{\nabla u|_{f_i}}{\|\nabla u|_{f_i}\|},
	\label{eq:HeatDiffusionNormalization}
\end{equation}
where $\nabla u|_{f_i}$ is the gradient of $u$ on face $f_i$.
The initial value of the scalar field $u$ is the Dirac delta function at the source vertex.
Integrating the heat flow using a single backward Euler step leads to a linear system~\cite{CraneWW13}:
\begin{equation}
  (\mathbf{A} - t \mathbf{L}_c) \mathbf{u} = \mathbf{u}_0,
  \label{eq:heat}
\end{equation}
where vector $\mathbf{u}$ stores the value of $u$ for each vertex, vector $\mathbf{u}_0$ has value 1 for the source vertex and 0 for all other vertices, $\mathbf{A}$ is a diagonal matrix storing the area of each vertex, and $\mathbf{L}_c$ is the cotangent Laplacian matrix. The solution to Eq.~\eqref{eq:heat} discretizes a scalar multiple of a function $v_t$ that is related to the geodesic distance $\phi$ from the source vertex via
\begin{equation}
	\lim_{t\rightarrow 0} - \frac{\sqrt{t}}{2} \log v_t = \phi
	\label{eq:HeatDistanceRelation}
\end{equation}
away from the cut locus~\cite{CraneWW13}. As the transformation to $v_t$ in Eq.~\eqref{eq:HeatDistanceRelation} does not alter its gradient direction, this relation ensures the validity of Eq.~\eqref{eq:HeatDiffusionNormalization} in approximating the gradient of the true geodesic distance.

In~\cite{CraneWW13}, this sparse linear system is solved by prefactorizing matrix $\mathbf{A} - t \mathbf{L}$ using Cholesky decomposition. This approach works well for meshes with up to a few million vertices, but faces difficulties for larger meshes due to high memory consumption for the factorization. Alternatively, we can solve the system using Krylov subspace methods such as conjugate gradient (CG), which requires less memory and can be parallelized~\cite{Greenbaum1989}. However, CG may require a large number of iterations to converge for large meshes, which still results in high computational costs.

\begin{figure}[t!]
\centering
\includegraphics[width=0.9\columnwidth]{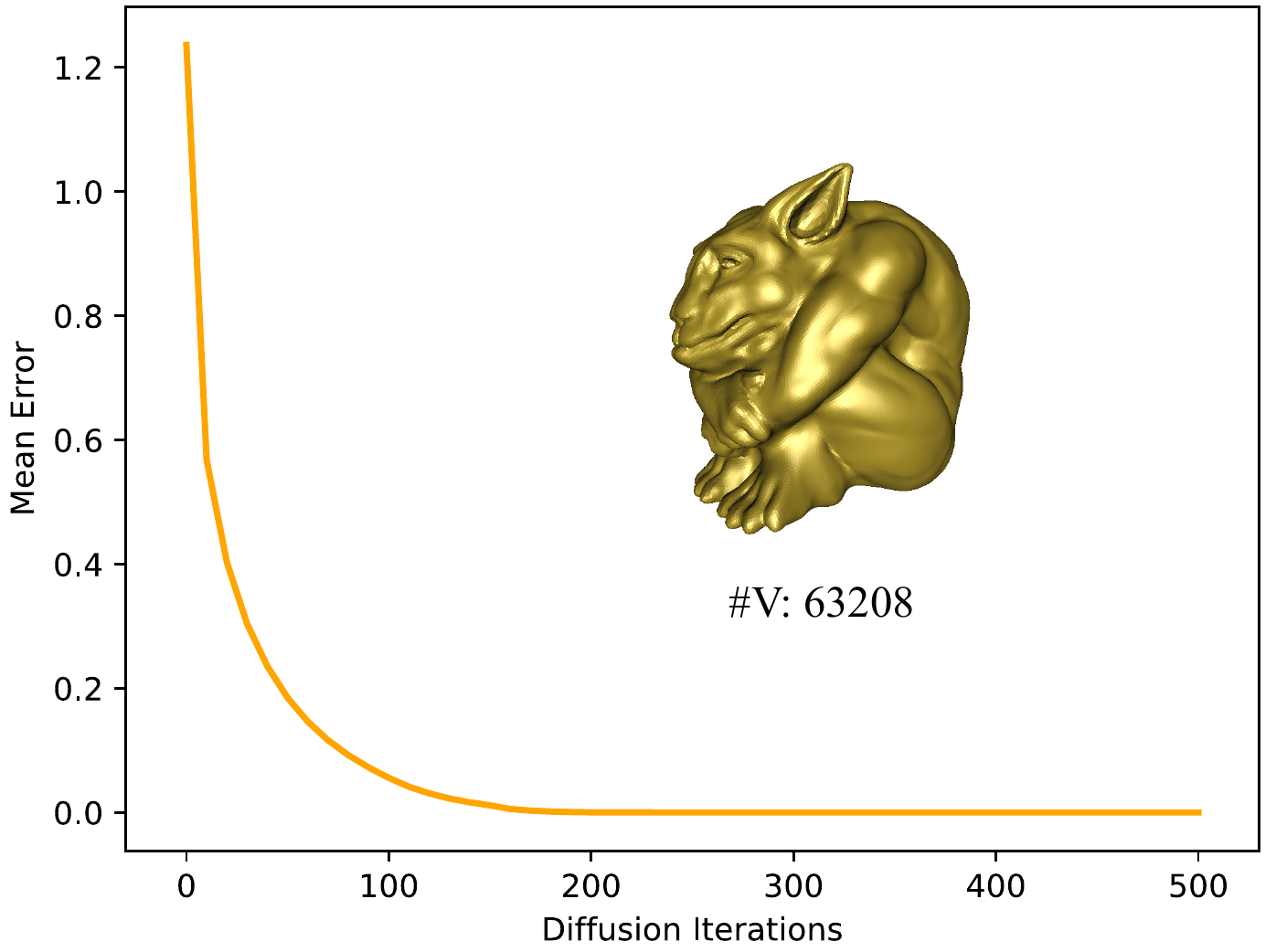}
\caption{Our Gauss-Seidel heat diffusion quickly decreases the mean error of the result, computed based on Eq.~\eqref{eq:relativeError}. As a result, it produces a solution good enough for subsequent steps with significantly less computational time than a direct solver or a Krylov subspace method.}
\label{fig:BFS-convergence}
\end{figure}

With scalability in mind, we prefer a method that can be run in parallel without requiring many iterations to converge. Therefore, we solve the system~\eqref{eq:heat} with Gauss-Seidel iteration in breadth-first order, where all vertices with the same topological distance to the source are updated in parallel. Specifically, let us define the sets
\begin{equation}
	\begin{aligned}
	& \mathcal{D}_0 \coloneqq \{v_s\}, \\
	& \mathcal{D}_1 \coloneqq \mathcal{N}(\mathbf{D}_0) \setminus \mathcal{D}_0, \\
	& \mathcal{D}_2 \coloneqq \mathcal{N}(\mathbf{D}_1) \setminus ( \mathcal{D}_0 \cup \mathcal{D}_1 ),\\
	& \ldots  \\
	& \mathcal{D}_i  \coloneqq  \mathcal{N}(\mathbf{D}_{i-1}) \setminus \bigcup_{k=0}^{i-1} \mathcal{D}_k,
	\end{aligned}
	\label{eq:VertexSets}
\end{equation}
where $\mathcal{N}(\cdot)$ denotes the union of one-ring neighbor vertices. Intuitively, ${D}_i$ is the set of vertices whose shortest paths to $v_s$ along mesh edges contain exactly $i - 1$ edges.  All such sets can be determined using breadth-first traversal of the vertices starting from $v_s$. Then in each outer iteration of our Gauss-Seidel solver, we update the values at vertex sets $\mathcal{D}_0, \mathcal{D}_1, \ldots$ consecutively, whereas the vertices belonging to the same set are updated simultaneously. When updating a set $\mathcal{D}_i$, we determine the new values for each vertex $v_j$ from the latest values of its neighboring vertices to satisfy its corresponding equation in~\eqref{eq:heat}, resulting in an update rule
\begin{equation}
	{u}^{i}(v_j)  = \frac{u_0(v_j) + t \sum_{k \in \mathcal{N}_j} \theta_{j,k} ~u^{i-1}(v_k)}
	{A_{v_j} + t \sum_{k \in \mathcal{N}(j)} \theta_{j,k}},
	\label{eq:HeatDiffusionUpdate}
\end{equation}
\begin{wrapfigure}{r}{0.5\columnwidth}
	\vspace*{-6pt}
	\centering
	\includegraphics[width=0.5\columnwidth]{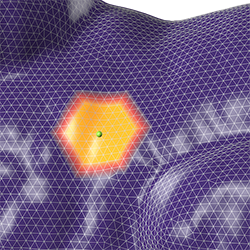}
	\vspace*{-14pt}
\end{wrapfigure}
where $A_{v_j}$ is the area for vertex $v_j$, computed as one-third of the total area of triangles incident with $v_j$; $\mathcal{N}_j$ denotes the index set of neighboring vertices for $v_j$, $u_0(v_j)$ is the initial value of $v_j$, $ u^{i-1}(v_k)$ is the value at $v_k$ after the update of $\mathcal{D}_{i-1}$, and $\theta_{j,k}$ is the coefficient of $v_k$ for the cotangent Laplacian at $v_j$. Intuitively, each outer iteration sweeps all vertices in breadth-first order starting from the source, where all vertices at the current front are updated simultaneously using the latest values from their neighbors. Note that unless it is close to the source, the set $\mathcal{D}_i$ often contains a large number of vertices. Thus the update can be easily parallelized with little overhead.
An illustration of our heat diffusion approach is shown in the inset. Here the green point is the source vertex, the yellow area covers vertices that have been updated in the current outer iteration, and the red front corresponds to vertices being updated in parallel. In our experiments, the solver quickly converges to a solution good enough for the subsequent steps, significantly reducing the computational time compared with the direct solver and Krylov subspace methods. An example is provided in Fig.~\ref{fig:BFS-convergence}, where we plot the following mean error $E$ of our solver after each outer iteration compared to the solution from the direct solver:
\begin{align}
E = \sqrt{ \sum\limits_{i=1}^{|\mathcal{F}|} A_i \| \mathbf{h}_i - \mathbf{h}^{*}_i\|^2 },
\label{eq:relativeError}
\end{align}
where $A_i$ is the area of face $f_i$ normalized by the total surface area, $\mathbf{h}_i$ and $\mathbf{h}^{*}_i$ are the normalized face gradient by our method and by directly solving Eq.~\eqref{eq:heat}, respectively. For this model with 63208 vertices, the direct solver takes 1.671s, while our method with four threads takes 153 outer iterations (0.169s) and 185 outer iterations (0.228s) to produce a result with mean error 1\% and 0.1\%, respectively.

\subsection{Integrable Gradient Field}
\label{sec:integrable}
In general, the unit vector field $\{\mathbf{h}_i\}$ is not integrable. To derive the geodesic distance, the heat method computes a scalar field whose gradients are as close as possible to $\{\mathbf{h}_i\}$, and shifts its values such that it vanishes at the source vertex. This is achieved in~\cite{CraneWW13} by solving a Poisson linear system
\begin{equation}
\mathbf{L} \mathbf{d} = \mathbf{b},
\end{equation}
where vector $\mathbf{d}$ stores the values of the scalar field, and $\mathbf{b}$ stores the integrated divergence of the field $\{\mathbf{h}_i\}$ at the vertices. Similar to heat diffusion, solving this linear system using Cholesky decomposition or Krylov subspace methods will face scalability issues. Therefore, we adopt a different strategy to derive geodesic distance: we first compute an integrable gradient field $\{\mathbf{g}_i\}$ that is the closest to $\{\mathbf{h}_i\}$, and then integrate it to recover the geodesic distance. In the following, we will show that both steps can be performed using efficient and scalable algorithms.

\begin{figure}[t!]
	\centering
	\includegraphics[width=0.5\columnwidth]{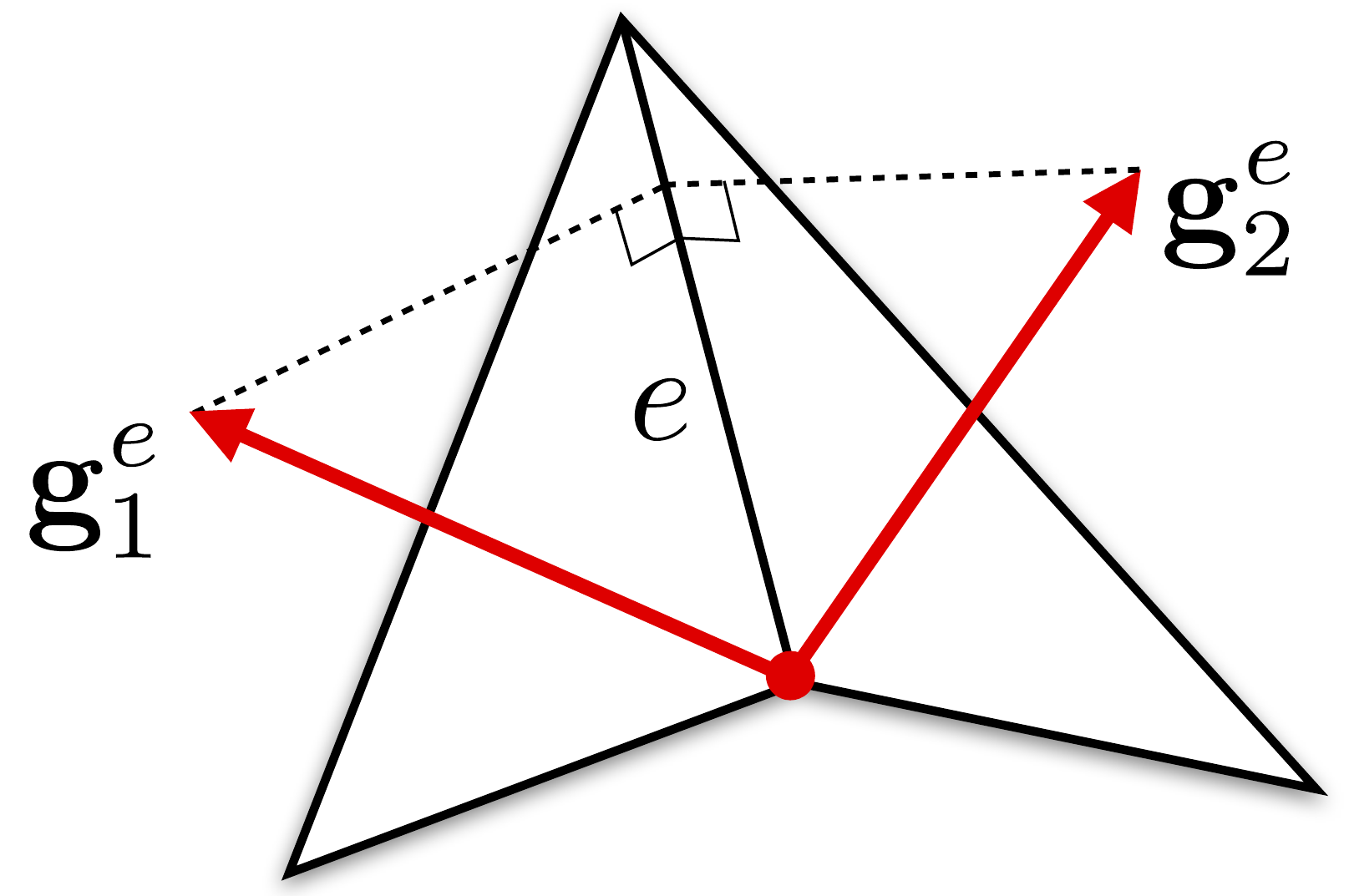}
	\caption{For a gradient field $\{\mathbf{g}_i\}$ to be integrable, the gradients $\mathbf{g}_1^e$, $\mathbf{g}_2^e$ on a pair of adjacent faces must have the same projection on their common edge $e$, resulting in the compatibility condition~\eqref{eq::GradCompatibility}.}
	\label{fig:CommonEdgeGradients}
\end{figure}

Representing $\{\mathbf{h}_i\}$ and $\{\mathbf{g}_i\}$ as 3D vectors, we compute $\{\mathbf{g}_i\}$ by solving a convex constrained optimization problem:
\begin{align}
\min_{\{\mathbf{g}_i\}} &\quad \sum_{f_i \in \mathcal{F}} A_i \|\mathbf{g}_i - \mathbf{h}_i\|^2 \label{eq:Integrable}, \\
\textrm{s.t.} &\quad \overline{\mathbf{e}} \cdot \left( \mathbf{g}_1^e - \mathbf{g}_2^e \right) = 0, ~~\forall~e \in \mathcal{E}_{\textrm{int}},
\label{eq::GradCompatibility}
\end{align}
where $\mathcal{E}_{\textrm{int}}$ denotes the set of interior edges, $\overline{\mathbf{e}} \in \mathbb{R}^3$ is the unit vector for edge $e$, and $ \mathbf{g}_1^e,  \mathbf{g}_2^e$ are the gradient vectors on the two incident faces for $e$.
Eq.~\eqref{eq::GradCompatibility} ensures the integrability of the gradient variables: the inner product between a face gradient and an incident edge vector yields the change of the underlying scalar function along the edge; therefore, if there exists a scalar function with the given face gradients, the gradient vectors on any pair of adjacent faces must have the same projection on their common edge (see Fig.~\ref{fig:CommonEdgeGradients}), which is equivalent to condition \eqref{eq::GradCompatibility}. 
To solve this problem in a scalable way, we introduce for each interior edge a pair of auxiliary variables $\mathbf{y}_1^e, \mathbf{y}_2^e \in \mathbb{R}^3$ for the gradients on its adjacent faces, and reformulate it as
\begin{align}
\min_{\{\mathbf{g}_i\}, \{(\mathbf{y}_1^e, \mathbf{y}_2^e)\}} &\quad \sum_{f_i \in \mathcal{F}} A_i \|\mathbf{g}_i - \mathbf{h}_i\|^2 + \sum_{e \in \mathcal{E}_{\textrm{int}}} \sigma_e(\mathbf{y}_1^e, \mathbf{y}_2^e) \label{eq:NewTarget}\\
\textrm{s.t.} &\quad \mathbf{g}_i = \mathbf{y}_k, ~~\forall f_i \in \mathcal{F},~\forall\mathbf{y}_k \in \mathcal{Y}_i. \label{eq:NewConstraint}
\end{align}
Here $A_i$ is the face area for $\mathbf{g}_i$, and $\sigma_e(\cdot)$ is an indicator function for compatibility between the auxiliary gradient variables on the two incident faces of edge $e$:
\begin{equation}
	\sigma_e(\mathbf{y}_1^e, \mathbf{y}_2^e)
	= \left\{
	\begin{array}{ll}
	0, & \textrm{if}~\overline{\mathbf{e}} \cdot (\mathbf{y}_1^e-\mathbf{y}_2^e) = 0, \\
	+ \infty, & \textrm{otherwise}.
	\end{array}
	\right.
\end{equation}
$\mathcal{Y}_i$ denotes the set of auxiliary variables associated with face $f_i$, such that constraint~\eqref{eq:NewConstraint} enforces consistency between the auxiliary variables and the actual gradient vectors. To facilitate presentation, we write it in matrix form as
\begin{align}
\min_{\mathbf{G}, \mathbf{Y}} &\quad \|\mathbf{M}(\mathbf{G} - \mathbf{H})\|_F^2 + \sigma(\mathbf{Y}),
\label{eq:ADMMMatrixFormTarget}\\
\textrm{s.t.} &\quad \mathbf{M} (\mathbf{Y} - \mathbf{S} \mathbf{G})  = \mathbf{0}.	
\label{eq:ADMMMatrixFormConstraint}
\end{align}
where $\mathbf{G}, \mathbf{H} \in \mathbb{R}^{|\mathcal{F}| \times 3}$ and $\mathbf{Y} \in \mathbb{R}^{ 2|\mathcal{E}_{\textrm{int}}| \times 3}$ collect the gradient variables, the input unit vector fields, and the auxiliary variables, respectively. $\sigma(\mathbf{Y})$ denotes the sum of indicator functions for all auxiliary variable pairs. $\mathbf{S} \in \mathbb{R}^{2|\mathcal{E}_{\textrm{int}}| \times |\mathcal{F}| }$ is a selection matrix that chooses the matching gradient variable for each auxiliary variable. $\mathbf{M} \in \mathbb{R}^{2|\mathcal{E}_{\textrm{int}}| \times 2|\mathcal{E}_{\textrm{int}}|}$ is a diagonal matrix storing the square roots of face areas associated with the auxiliary variables. Introducing $\mathbf{M}$ in the constraints does not alter the solution, but helps to make the algorithm robust to the mesh discretization. Indeed, a similar constraint reweighting strategy is employed in~\cite{Overby2017} to improve the convergence of their ADMM solver for physics simulation.

To solve this problem, ADMM searches for a saddle point of its augmented Lagrangian
\begin{align*}
    L(\mathbf{G}, \mathbf{Y}, \boldsymbol{\lambda}) = & \|\mathbf{M}(\mathbf{G} - \mathbf{H})\|_F^2 + \sigma(\mathbf{Y})\\
    & + \Tr \left(\boldsymbol{\lambda}^T   \mathbf{M} (\mathbf{Y} - \mathbf{G} \mathbf{S})\right)
    + \frac{\mu}{2} \| \mathbf{M} (\mathbf{Y} - \mathbf{S} \mathbf{G} ) \|_F^2,
\end{align*}
where $\boldsymbol{\lambda} \in \mathbb{R}^{2 |\mathcal{E}_{\textrm{int}}| \times 3}$ collects the dual variables, and $\mu > 0$ is a penalty parameter. We choose $\mu = 100$ in this paper. The stationary point is computed by alternating between the updates of $\mathbf{Y}$, $\mathbf{G}$, and $\boldsymbol{\lambda}$.

\begin{itemize}
	\item \textbf{$\mathbf{Y}$-update.} We minimize the augmented Lagrangian with respect to $\mathbf{Y}$ while fixing $\mathbf{G}$ and $\boldsymbol{\lambda}$, i.e.,
\[
\min_{\mathbf{Y}} ~~\sigma(\mathbf{Y}) + \frac{\mu}{2}\left\|\mathbf{M}(\mathbf{Y} - \mathbf{S}\mathbf{G}) + \frac{\boldsymbol{\lambda}}{\mu}\right\|_F^2.
\]
This is separable into a set of independent subproblems for each internal edge $e$:
\begin{displaymath}
	\min_{\mathbf{y}_1^e, \mathbf{y}_2^e}~~\sigma_e(\mathbf{y}_1^e, \mathbf{y}_2^e) + \frac{\mu}{2}\sum_{i=1}^2 \left\|\alpha_i^e (\mathbf{y}_i^e - \mathbf{g}_i^e) + \frac{\boldsymbol{\lambda}_i^e}{\mu}\right\|^2,
\end{displaymath}
where $\alpha_i^e$, $\mathbf{g}_i^e$, $\boldsymbol{\lambda}_i^e$ ($i=1,2$) are the face area square root, gradient variables, and dual variables corresponding to $\mathbf{y}_i^e$, respectively. This problem has a closed-form solution
\begin{eqnarray*}
	\mathbf{y}_1^e &=& \mathbf{q}_1^e + \frac{A_2^e}{A_1^e + A_2^e}\overline{\mathbf{e}} \left(\overline{\mathbf{e}} \cdot (\mathbf{q}_2^e - \mathbf{q}_1^e)\right), \\ 
	\mathbf{y}_2^e &=& \mathbf{q}_2^e - \frac{A_1^e}{A_1^e + A_2^e}\overline{\mathbf{e}} \left(\overline{\mathbf{e}} \cdot (\mathbf{q}_2^e - \mathbf{q}_1^e)\right),
\end{eqnarray*}
where $\mathbf{q}_i^e = \mathbf{g}_i^e - \frac{\boldsymbol{\lambda}_i^e}{\mu \alpha_i^e}$, and $A_i^e$ is the face area for $\mathbf{y}_i^e$, for $i=1,2$.

\item \textbf{$\mathbf{G}$-update.} After updating $\mathbf{Y}$, we fix $\mathbf{Y}, \boldsymbol{\lambda}$ and minimizes the augmented Lagrangian with respect to $\mathbf{G}$, which reduces to independent subproblems
\begin{align}
\min_{\mathbf{g}_i}~~ A_i \|\mathbf{g}_i - \mathbf{h}_i\|^2 + \frac{\mu}{2} \sum_{\mathbf{y}_k \in \mathcal{Y}(i)} \left\| \alpha_i( \mathbf{y}_k - \mathbf{g}_i ) + \frac{\boldsymbol{\lambda}_k}{\mu } \right\|^2,
\label{eq:G-subproblem}
\end{align}
where $\mathcal{Y}(i)$ denotes the set of associated auxiliary variables for ${\mathbf{g}_i}$ in $\mathbf{Y}$, and $\boldsymbol{\lambda}_k$ denotes the corresponding components in $\boldsymbol{\lambda}$ for $\mathbf{y}_k$. These subproblems can be solved in parallel with a closed-form solution
\[
\mathbf{g}_i = \frac{2 \mathbf{h}_i + \mu \sum_{\mathbf{y}_k \in \mathcal{Y}(i)}(\mathbf{y}_k + \frac{\boldsymbol{\lambda}_k}{\mu\alpha_i})}{2 + \mu |\mathcal{Y}(i)|}.
\]

\item \textbf{$\boldsymbol{\lambda}$-update.}
After the updates for $\mathbf{Y}$ and $\mathbf{G}$, we compute the new values $\boldsymbol{\lambda'}$ for the dual variables as
\begin{displaymath}
	\boldsymbol{\lambda'} = \boldsymbol{\lambda} + \mu \mathbf{M} (\mathbf{Y} - \mathbf{S} \mathbf{G}).
\end{displaymath}
\end{itemize}
To initialize the solver, on each face $f_i$ we set the gradient variable $\mathbf{g}_i$ and each auxiliary variable $\mathbf{y}_k \in \mathcal{Y}_i$ to $\mathbf{h}_i$, whereas all dual variables $\boldsymbol{\lambda}$ are set to zero.
Since our optimization problem is convex, ADMM converges to a stationary point of the problem~\cite{Boyd2011-admm}. We measure the convergence using the primal residual $\mathbf{r}_{\textrm{primal}}$ and the dual residual $\mathbf{r}_{\textrm{dual}}$~\cite{Boyd2011-admm}:
\[
	\mathbf{r}_{\textrm{primal}} =
	\mathbf{M}(\mathbf{Y} - \mathbf{S} \mathbf{G}),
	\quad
	\mathbf{r}_{\textrm{dual}} = 
	\mu \mathbf{A} \mathbf{S} \delta \mathbf{G},
\]
where $\delta \mathbf{G}$ is the difference of $\mathbf{G}$ between two iterations.
We terminate the algorithm when both of the primal residual and dual residual are small enough,
or if the iteration count exceeds a user-specified threshold $I_{\max{}}$.
The residual thresholds are set to $\|\mathbf{r}_{\textrm{primal}}\| \leq \|\mathbf{M}\|_F \cdot \epsilon_1$ and $\|\mathbf{r}_{\textrm{dual}}\| \leq \|\mathbf{M}\|_F \cdot \epsilon_2$, where $\epsilon_1, \epsilon_2$ are user-specified values. We set $\epsilon_1 = \epsilon_2 = 1 \cdot 10^{-5}$ in all our experiments.

\begin{figure}[t!]
	\centering
	\includegraphics[width=0.9\columnwidth]{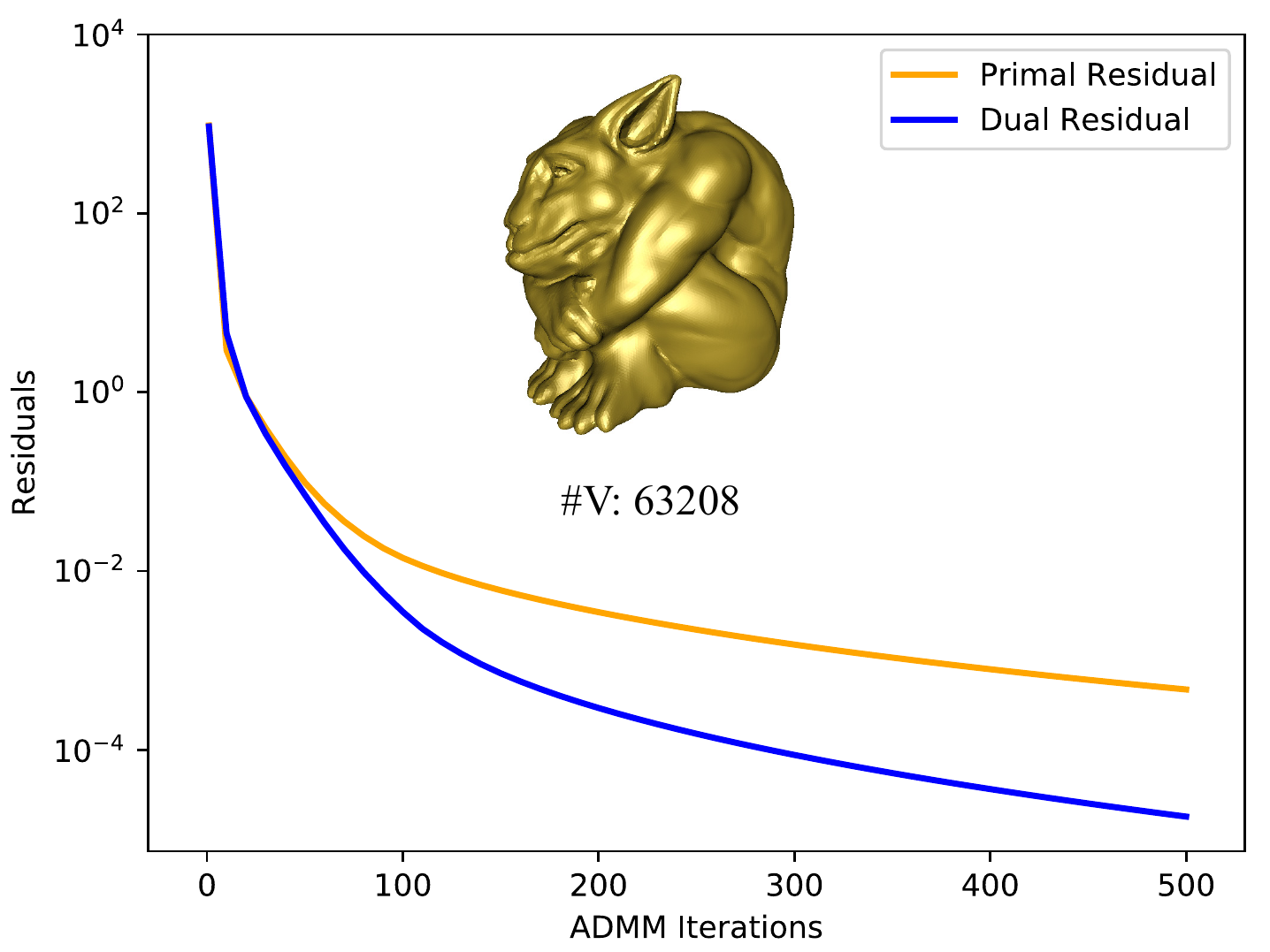}
	\caption{The ADMM solver quickly decreases the primal and dual residuals in the initial iterations, as shown here on the same model as in Fig.~\ref{fig:BFS-convergence}.}
	\label{fig:ADMM-convergence}
\end{figure}

\begin{figure*}[t!]
	\centering
	\includegraphics[width=\textwidth]{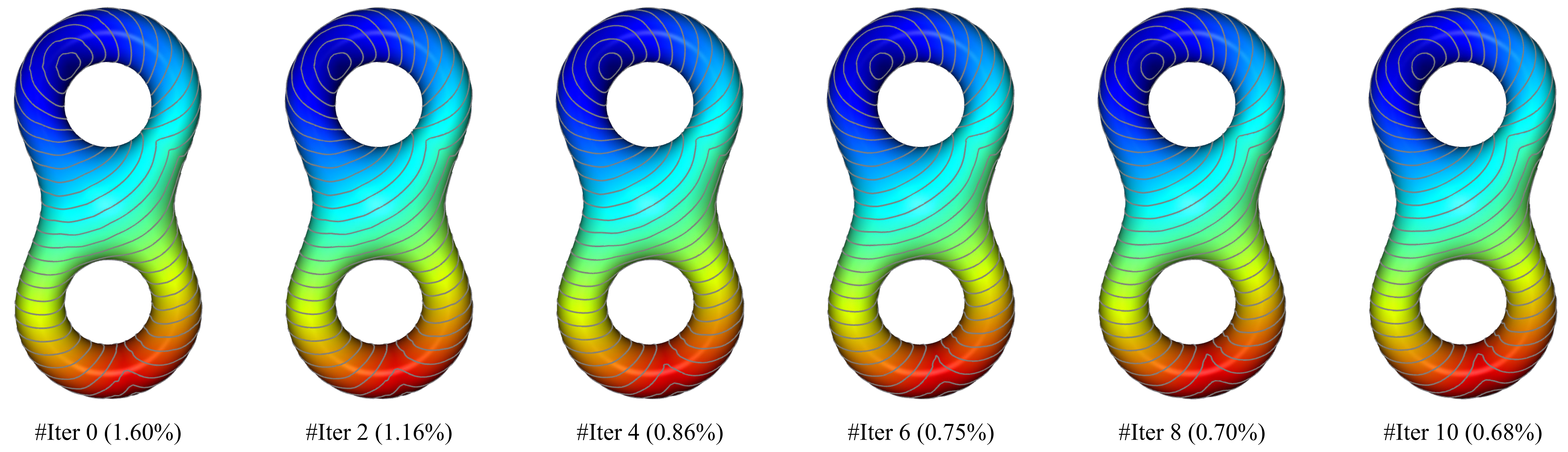}
	\caption{Starting from the unit gradients resulting from heat diffusion, our ADMM solver can produce a geodesic distance gradient field of good accuracy within a small number of iterations. Here we run the ADMM solver for a prescribed number of iterations, and integrate the resulting gradients according to Section~\ref{sec:integration}. The recovered geodesic distance field is visualized with color coding and level sets. The caption of each image shows the number of ADMM iterations and the mean relative error of the resulting geodesic distance, computed according to Eq.~\eqref{eq:MeanRelError}. }
	\label{fig:pipeline}
\end{figure*}

The main benefit of ADMM is its fast convergence to a point close to the final solution~\cite{Boyd2011-admm}. In our experiments, the algorithm only needs a small number of iterations to reduce both the primal and dual residuals to small values, as shown in Fig.~\ref{fig:ADMM-convergence}). As a result, it only takes a small number of iterations for the ADMM solver to produce a gradient field with good accuracy compared to the exact solution (see Fig.~\ref{fig:pipeline} for an example).
Moreover, the updates of $\mathbf{Y}$, $\mathbf{G}$, and $\boldsymbol{\lambda}$ are all trivially parallelizable, allowing for significant speedup on multi-core processors. Finally, the memory consumption grows linearly with the mesh size, making it feasible to process very large meshes. Therefore, our method is both efficient and scalable.

\noindent\textbf{Remark.} Our formulation using gradients as variables is a key factor in achieving efficiency and scalability for ADMM. In the past, ADMM and other first-order methods have been used to solve optimization problems that regularize the gradient of certain functions~\cite{WangYYZ08,NgWY10,XuLXJ11,HeideDNRHW16}. These problems are all formulated with the function values as variables. For such problems, the solver typically involves a local step that updates auxiliary gradient variables according to the regularization, and a global step that updates the function variables to align with the auxiliary gradients. The global step requires solving a linear system for all the function variables, which will eventually become a bottleneck for large-scale problems. By formulating the problem with gradient variables instead, our global step reduces to a simple weighted averaging of a few auxiliary gradients, which is separable between different faces and can be done in parallel. From another perspective, for formulations that use function values as variables, the global step integrates the auxiliary gradients, which is globally coupled and limits parallelism. By using gradient variables, we bypass this time-consuming global integration step, and postpone it to a later stage where the optimized gradients are integrated once to recover the geodesic distance.

\subsection{Integration}
\label{sec:integration}
After computing an integrable gradient field $\mathbf{g}_i$, we determine the geodesic distance $d$ at each vertex by setting $d(v_s) = 0$ and integrating $\mathbf{g}_i$ starting from the source $v_s$. Similar to the heat diffusion process, we determine the geodesic distance in breadth-first order, processing the vertex sets $\mathcal{D}_1, \mathcal{D}_2, \ldots$ consecutively. For a vertex $v_j \in \mathcal{D}_{i}$ ($i \geq 0$), its geodesic distance is determined from a neighboring vertex $v_k \in \mathcal{D}_{i-1}$ via
\begin{equation}
	d(v_j) = d(v_k) + \frac{1}{|\mathcal{T}_{jk}|} \sum_{f_l \in \mathcal{T}_{jk}} \mathbf{g}_l \cdot (\mathbf{p}_j - \mathbf{p}_k), 
	\label{eq:IntegrationStep}
\end{equation}
where $\mathbf{p}_j,\mathbf{p}_k \in \mathbb{R}^3$ are the positions of $v_j, v_k$, and $\mathcal{T}_{jk}$ denotes the set of faces that contain both $v_j$ and $v_k$. The pairing between $v_j$ and $v_k$ can be determined using breadth-first traversal from the source vertex. In our implementation, we pre-compute the vertex sets $\{\mathcal{D}_i\}$ as well as the vertex pairing using one run of breadth-first traversal, and reuse the information in the heat diffusion and geodesic distance integration steps. Like our heat diffusion solver, the integration step~\eqref{eq:IntegrationStep} is independent between the different vertices within $\mathcal{D}_{i}$, and can be performed in parallel with little overhead.

\section{Edge-based Optimization}
\label{sec:edgebased}

For the optimization formulation in Section~\ref{sec:algorithm}, the gradient variables and auxiliary variables are 3D vectors defined on faces. This can lead to redundant memory storage: since a gradient vector must be orthogonal to the face normal, it has only two degrees of freedom. This intrinsic property is disregarded by the 3D vector representation that encodes the gradients with respect to the ambient space. To reduce  memory footprint, we may encode a gradient vector and its associated auxiliary variables as 2D local coordinates its the face. However, this would make the compatibility conditions between two neighboring gradients more complicated, as we must first transform them into a common frame. This not only makes the update of auxiliary variables more involved, but also requires additional storage for the transformation. 

In this section, we propose a new formulation for the gradient optimization problem that comes with a smaller memory footprint while maintaining the simplicity of the solver steps. The key idea is that if a face-based gradient $\mathbf{g}$ is the same as its target value $\mathbf{h}$, then their projections onto the three triangle edges must be the same as well. Moreover, the inner product between the gradient $\mathbf{g}$ and an edge vector $\mathbf{e} = \mathbf{v}_1 - \mathbf{v}_2$ encodes the change of the underlying scalar function $d$ between its two vertices ${v}_1$ and ${v}_2$, i.e.,
$\mathbf{g} \cdot \mathbf{e}  = d({v}_1) - d({v}_2).$
Furthermore, given a non-degenerate triangle, any vector orthogonal to its normal can be uniquely recovered from the inner products between the vector and its three edge vectors. Therefore, we can encode the geodesic distance gradients $\{\mathbf{g}_i\}$ and their targets $\{\mathbf{h}_i\}$ using their inner products with their respective triangle edges. And instead of penalizing the difference between $\mathbf{g}_i$ and $\mathbf{h}_i$ directly, we can penalize the difference between their inner products with edge vectors.

In the following, we first present the optimization formulation and its solver based on this idea. Afterward, we analyze its memory consumption in Section~\ref{sec:EdgeBasedAnalysis}, to show that it is indeed more memory-efficient than the face-based optimization approach.

\subsection{Formulation}
\label{sec:edgeformulation}
\begin{figure}[t!]
	\centering
	\includegraphics[width=0.5\columnwidth]{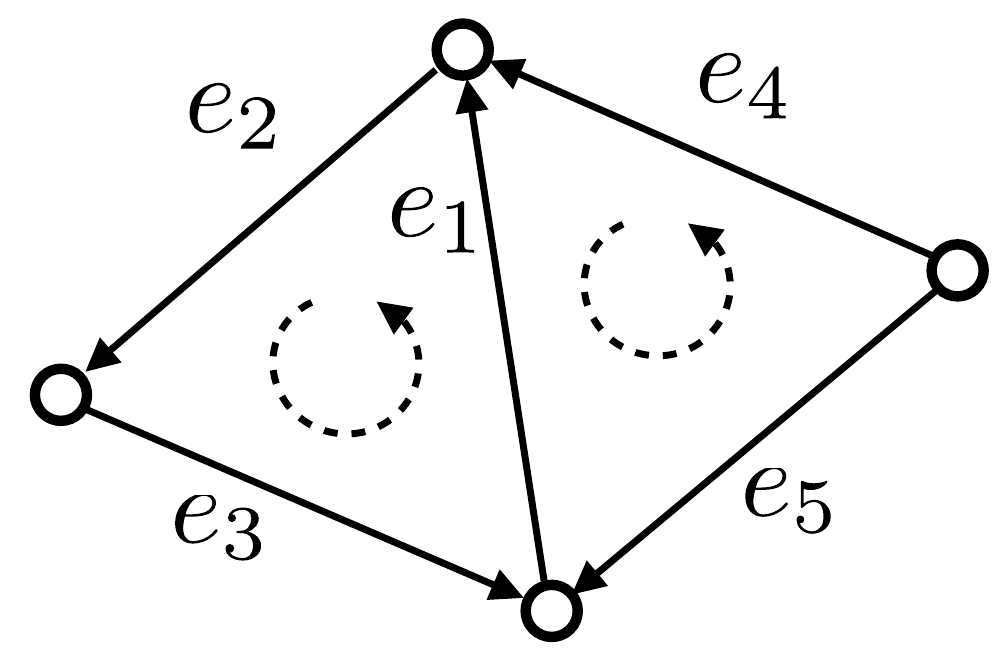}
	\caption{The integrability condition for the edge variables within a triangle depends on the direction of their orientation halfedges, shown using arrows in this figure. The dashed arcs show the orientation for each triangle. For this particular example, the integrability conditions are $x_{e_1} + x_{e_2} + x_{e_3} = 0$ and $-x_{e_1} + x_{e_4} - x_{e_5} = 0$, respectively.}
	\label{fig:EdgeIntegrability}
\end{figure}

To compute a geodesic distance function $d$ on a triangle mesh, we define for each edge $e$ a scalar variable $x_e$ that represents the difference of $d$ on its two vertices. It corresponds to the change of $d$ along one of the halfedges of $e$. We call this halfedge its \emph{orientation halfedge} $\eta_e$.
As mentioned previously, $x_e = \mathbf{g} \cdot \mathbf{e}$, where $\mathbf{g}$ is the gradient of $d$ on an incident faces $f$ of $e$, and $\mathbf{e}$ is the vector for the orientation halfedge $\eta_e$. Let $\mathbf{h}$ be the target normalized gradient on $f$, and denote 
\begin{equation}
	h_e = \mathbf{h} \cdot \mathbf{e}.
	\label{eq:TargetDifference}
\end{equation}
Then we can penalize the deviation between $\mathbf{g}$ and $\mathbf{h}$ using their squared difference along edge $e$, resulting in an error term $(x_e - h_e)^2$.
Thus we compute the geodesic distance function across the whole mesh via an optimization problem
\begin{eqnarray}
	\min_{\mathbf{X}} && \frac{1}{2}  \sum_{e \in \mathcal{E}} \sum_{\mathbf{h} \in \mathcal{H}_e} (x_e - h_e)^2,\\
	\textrm{s.t.} && \sum_{k=1}^3 s_k^f \cdot x_{e_k^f} = 0,~~ \forall f \in \mathcal{F}. \label{eq:Integrability}
\end{eqnarray}  
Here $\mathcal{E}$ is the set of mesh edges; $\mathcal{F}$ is the set of mesh faces; $\mathbf{X} \in \mathbb{R}^{|\mathcal{E}|}$ collects the edge variables;  $\mathcal{H}_e$ denotes the set of target gradients on incident faces of $e$, such that $|\mathcal{H}_e|$ is either 1 or 2 for any edge that is not isolated. The constraint~\eqref{eq:Integrability} is an integrability condition for the variables $\{x_e\}$, so that the total change of function $d$ along an edge loop must vanish. Here $e_1^f, e_2^f, e_3^f$ are the three edges incident with a face $f$, and $s_1^f, s_2^f, s_3^f \in \{-1, 1\}$ are their signs relative to the edge loop of $f$ (see Fig.~\ref{fig:EdgeIntegrability} for an example). Note that for orientable 2-manifold meshes represented using halfedge data structures~\cite{Campagna1998,Kettner1999}, the halfedges incident with a face must form an oriented loop. and the value of $s_k^f$ ($k=1,2,3$) can be easily determined from the orientation halfedge $\eta_{e_k^f}$: $s_k^f = 1$ if $\eta_{e_k^f}$ is incident with face $f$, otherwise $s_k^f=-1$.

To solve this optimization problem, we first introduce for each face $f$ three auxiliary scalar variables $w_{e_1^f}, w_{e_2^f}, w_{e_3^f}$ corresponding to the edge variables $x_{e_1^f}, x_{e_2^f}, x_{e_3^f}$. Then we can reformulate the problem in a matrix form as
\begin{eqnarray*}
	\min_{\mathbf{X}, \mathbf{W}} && \frac{1}{2} \|\mathbf{R} \mathbf{X} - \mathbf{Z}\|^2 + \sigma(\mathbf{W}),\\
	\textrm{s.t.} && \mathbf{W} - \mathbf{R} \mathbf{X} = 0.
\end{eqnarray*}
Here $\mathbf{X}$ collects all edge variables $\{x_e\}$; $\mathbf{Z} \in \mathbb{R}^{3|\mathcal{F}|}$ collects all target difference values $\{h_e\}$, arranged in triplets each induced by the target gradient on a face; $\mathbf{R} \in \mathbb{R}^{3|\mathcal{F}| \times |\mathcal{E}|}$ is a sparse selection matrix that chooses the edge variables according to its target difference values; $\mathbf{W} \in \mathbb{R}^{3 |\mathcal{F}|}$ collects the auxiliary variables arranged in the same order as $\mathbf{Z}$; $\sigma$ is an indicator function for the integrability condition~\eqref{eq:Integrability}:
\[
\sigma(\mathbf{W}) = \left\{
	\begin{array}{ll}
	0 & \textrm{if}~\sum_{k=1}^3 s_k^f \cdot w_{e_k^f} = 0~~ \forall f \in \mathcal{F},\\
	+\infty & \textrm{otherwise}.
	\end{array}
\right.
\]
Similar to the face-based method, we solve this problem using ADMM.
Its augmented Lagrangian function is
\begin{align*}
L(\mathbf{X}, \mathbf{W}, \boldsymbol{\lambda})
= &~ \frac{1}{2} \|\mathbf{Z} - \mathbf{R} \mathbf{X}\|^2 + \sigma (\mathbf{W}) \\
& + \boldsymbol{\lambda} \cdot  (\mathbf{W} - \mathbf{R} \mathbf{X}) + \frac{\mu}{2} \|\mathbf{W} - \mathbf{R} \mathbf{X}\|^2,
\end{align*}
where $\boldsymbol{\lambda} \in \mathbb{R}^{3 |\mathcal{F}|}$ and $\mu > 0$ are the dual variables and the penalty parameter, respectively. We choose $\mu = 100$ in this paper.
ADMM finds a stationary point of $L(\mathbf{X}, \mathbf{W}, \boldsymbol{\lambda})$ by alternating between three steps:
\begin{itemize}
	\item \textbf{Fix $\mathbf{X}, \boldsymbol{\lambda}$, update $\mathbf{W}$.} This reduces to separable subproblems for each face $f$:
	\begin{eqnarray*}
	\min_{\mathbf{w}_f} && \left\|\mathbf{w}_f - \mathbf{x}_f + \frac{\boldsymbol{\lambda}_f}{\mu} \right\|^2 \\
	\textrm{s.t.} && \mathbf{q}_f \cdot \mathbf{w}_f  = 0,
	\end{eqnarray*}
	where $\mathbf{w}_f \in \mathbb{R}^3$ are the components of $\mathbf{W}$ corresponding to face $f$, $\mathbf{q}_f \in \{1, -1\}^3$ stores the signs for the components of $\mathbf{w}_f$ in the integrability condition,
	and $\mathbf{x}_f, \boldsymbol{\lambda}_f$ are the corresponding components of $\mathbf{X}$ and $\boldsymbol{\lambda}$ respectively.
	The problem has a closed-form solution
	\[
		\mathbf{w}_f = (\mathbf{I}_3 - \frac{1}{3} \mathbf{q}_f \mathbf{q}_f^T) (\mathbf{x}_f - \frac{\boldsymbol{\lambda}_f}{\mu}),
	\]
	where $\mathbf{I}_3$ is the $3 \times 3$ identity matrix.
	
	\item \textbf{Fix $\mathbf{W}, \boldsymbol{\lambda}$, update $\mathbf{X}$.} This leads to independent subproblems for each edge $e$:
	\[
	\min_{x_e} ~~ \sum_{\mathbf{h} \in \mathcal{H}_e}(x_e - h_e)^2 + \mu (w^e_{\mathbf{h}} - x_e + \frac{\lambda^e_{\mathbf{h}}}{\mu})^2,
	\]
	where ${w}^e_{\mathbf{h}}$ is the component of $\mathbf{W}$ for edge $e$ induced by a target gradient $\mathbf{h}$, and $\lambda^e_{\mathbf{h}}$ is the corresponding component of $\boldsymbol{\lambda}$. It has closed-form solution
	\begin{equation}
		x_e = \frac{\sum_{\mathbf{h} \in \mathcal{H}_e} (h_e + \lambda^e_{\mathbf{h}} + \mu w^e_{\mathbf{h}})}{|\mathcal{H}_e|(1 + \mu)}.
		\label{eq:xeupdate}
	\end{equation}
	
	\item \textbf{Fix $\mathbf{X}, \mathbf{W}$, update $\boldsymbol{\lambda}$}. The updated dual variables $\boldsymbol{\lambda'}$ are computed as
	\[
	\boldsymbol{\lambda'} = \boldsymbol{\lambda} +  \mu (\mathbf{W} - \mathbf{R} \mathbf{X}).
	\]
\end{itemize}
To initialize the solver, $\boldsymbol{\lambda}$ is set to zero, each edge variable $x_e$ is set to the average of its associated target values, and the three auxiliary variables on each face $f$ are set to the target values induced on its three edges by the target gradient on $f$.
The convergence of the solver is indicated by small norms of the primal and dual residuals:
\[
\mathbf{r}_{\textrm{primal}} = \mathbf{W} - \mathbf{R} \mathbf{X},
\quad
\mathbf{r}_{\textrm{dual}} = 
\mu \mathbf{R} \delta \mathbf{X},
\]
where $\delta \mathbf{X}$ is the difference of $\mathbf{X}$ in two consecutive iterations.

\subsection{Analysis}
\label{sec:EdgeBasedAnalysis}
Compared with the face-based formulation in Section~\ref{sec:integrable}, our edge-based formulation has a much lower memory footprint.
Let $N_E$ and $N_F$ be the number of vertices, edges, and faces in the mesh, respectively. The face-based formulation requires at least the following storage:
\begin{itemize}
	\item For each face: two 3D vectors for the gradient variable and the target gradient, respectively.
	\item For each internal edge: two 3D vectors for the auxiliary gradient variables, two 3D vectors for the dual variables, and one 3D vector for the edge direction used in the $\mathbf{Y}$-update.
\end{itemize}
Therefore, for a mesh with no boundary, the face-based formulation requires at least a storage space of $(6 N_F + 15 N_E) \cdot M $ bytes, where $M$ is the width of a floating point value. Accordingly, the edge-based formulation requires the following storage:
\begin{itemize}
	\item For each edge: one scalar for the variable $x_e$, and one scalar for a pre-computed value $\sum_{\mathbf{h} \in \mathcal{H}_e} h_e$ used in the update equation~\eqref{eq:xeupdate}.
	\item For each face: three scalars for auxiliary variables, three flags of their signs in the integrability condition, and three scalars for the dual variables.
\end{itemize}
Thus the edge-based formulation requires storage of $(6 N_F + 2 N_E) \cdot M + 3 N_F \cdot K $ bytes, where $K$ is the width of a sign flag. As the sign only takes two possible values, we can store it with as little storage as a single bit. Even if we store it using an integral data type, we can still ensure $K \leq M / 2$. Moreover, for a mesh without boundary, we have $N_E = \frac{3}{2} N_F$. Thus the edge-based formulation can at least reduce the storage by $12 N_E \cdot M$ bytes, which can be significant for large models. In our experiments, the edge-based solver can reduce the peak memory consumption by about 50\% compared to the face-based solver. A detailed comparison is provided in Table~\ref{tab:results}.

\noindent\textbf{Remark.} There is an interesting interpretation of our edge-based formulation from the perspective of discrete exterior calculus~\cite{Crane:2013:DGP,Desbrun2008-DEC}. The edge difference variables, being scalars defined on the edges, can be considered as a \emph{discrete 1-form}. If the edge difference variables satisfy the condition~\eqref{eq:Integrability}, then they are a \emph{closed 1-form}. On a simply-connected mesh surface (i.e., a topological disk), any closed 1-form is also an \emph{exact 1-form}, meaning that it corresponds to the derivative of a scalar function. Thus condition~\eqref{eq:Integrability} ensures the integrability of our edge difference variables. This relationship also holds in the smoothing setting: any closed 1-form on a simply-connected domain is also exact. This may help to generalize our formulation to other domains without a triangulated structure such as point clouds.
\section{Extensions}
\label{sec:extensions}

Our methods can be extended to handle surfaces with complex topologies, multiple sources, and poor triangulation. In this paper, these extensions are only applied to the examples shown in this section. Moreover, in practice our methods work well for complex topologies and multiple sources even without the extensions.

\subsection{Complex Topology}
\begin{wrapfigure}{r}{0.26\columnwidth}
	\vspace*{-6pt}
	\centering
	\includegraphics[width=0.26\columnwidth]{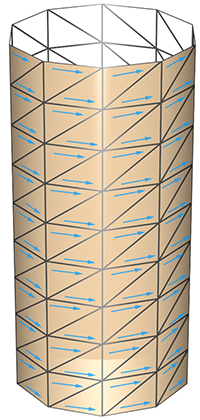}
	\vspace*{-20pt}
\end{wrapfigure} 
On a mesh surface of genus zero, the compatibility condition~\eqref{eq::GradCompatibility} ensures that the face gradients are globally integrable. This is no longer true for other topologies. 
One example is shown in the inset.
For this cylindrical mesh we can construct a unit vector field that is consistently oriented along the directrix and satisfies the compatibility condition~\eqref{eq::GradCompatibility}. This vector field is not the gradient field of a scalar function, however, because integrating along a directrix will result in a different value when coming back to the starting point. In general, for a tangent vector field on a surface of arbitrary topology to be a gradient field, we must ensure its line integral along any closed curve vanishes. On a mesh surface of genus $g$, such a closed curve can be generated from a cycle basis that consists of $2g$ independent non-contractible cycles~\cite{Crane2010}. Therefore, in addition to the compatibility condition~\eqref{eq::GradCompatibility}, we enforce an integrability condition of the field $\{\mathbf{g}_i\}$ on each cycle in the basis. 

\begin{wrapfigure}{r}{0.4\columnwidth}
	\vspace*{-6pt}
	\centering
		\includegraphics[width=0.4\columnwidth]{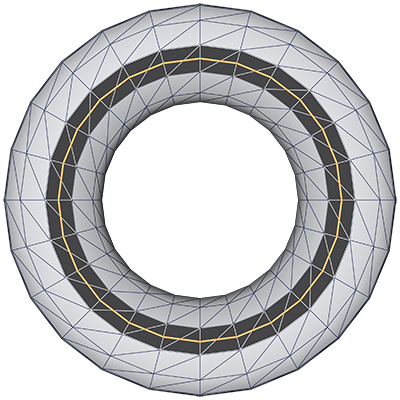}
	\vspace*{-12pt}
\end{wrapfigure}
Specifically, for the face-based method, we follow~\cite{Crane2010} and compute a cycle basis for the dual graph of the mesh, using the tree-cotree decomposition from~\cite{Eppstein2003}. 
Each cycle is a closed loop $C$ of faces. Connecting the mid-points of edges that correspond to the dual edges on $C$, we obtain a closed polyline $P$ lying on the mesh surface, with each segment lying on one face of the loop (see inset).  
Then the integrability condition on this cycle can be written as
\begin{equation}
	\sum_{f_i \in \mathcal{C}_F} \mathbf{g}_i \cdot \mathbf{v}_i^P = 0,
	\label{eq:TopologyConstraint}
\end{equation}
where $\mathcal{C}_F$ is the set of faces on the cycle, and $\mathbf{v}_i^P \in \mathbb{R}^3$ is the vector of polyline segment on face $f_i$ in the same orientation as the cycle. 
Adding these conditions to the optimization problem~\eqref{eq:Integrable}-\eqref{eq::GradCompatibility}, we derive a new formulation that finds a globally integrable gradient field closest to $\{\mathbf{h}_i\}$. 

For the edge-based method, we compute a cycle basis on the mesh itself, with each cycle being a closed loop of edges. Then for each such edge cycle we add the following constraint to the optimization problem:
\begin{equation}
	\sum_{e_i \in \mathcal{C}_E} s_i^{\mathcal{C}_E} \cdot x_{e_i} = 0,
	\label{eq:EdgeTopologyConstraint}
\end{equation}
where $\mathcal{C}_E$ is the set of edges on the cycle, and $s_i^{\mathcal{C}_E} \in \{-1, 1\}$ indicates the orientation of edge $e_i$ with respect to the cycle.

\begin{figure}[t]
	\centering
	\includegraphics[width=\columnwidth]{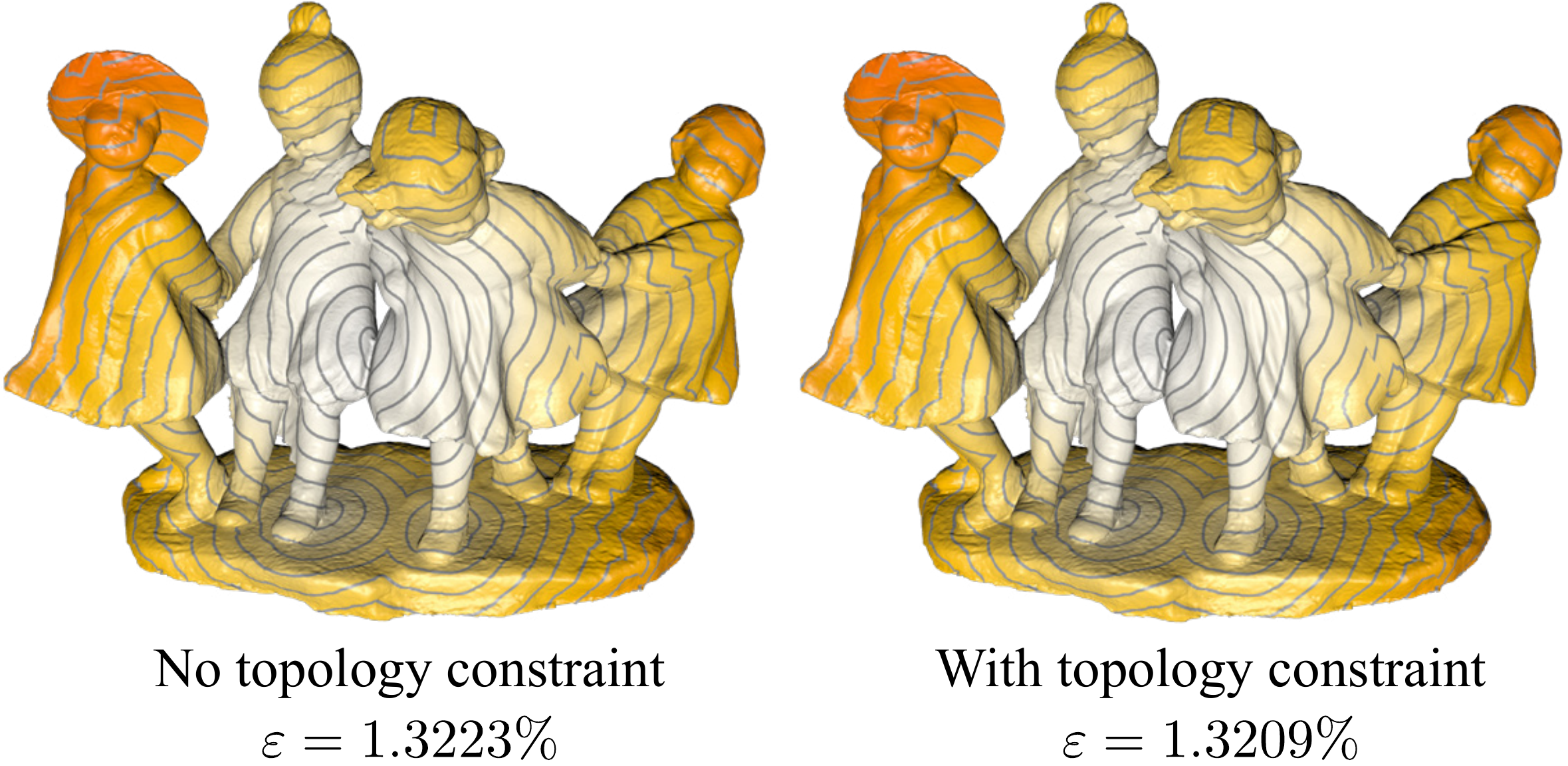}
	\caption{Geodesic distance on a surface of genus eight, computed using our face-based method with and without the constraint~\eqref{eq:TopologyConstraint}, and its mean relative error $\varepsilon$ compared to the ground truth as defined in Eq.~\eqref{eq:MeanRelError}. }
	\label{fig:topology}
\end{figure} 

After adding the constraints~\eqref{eq:TopologyConstraint} or~\eqref{eq:EdgeTopologyConstraint}, the new optimization problem is still convex and can be solved using ADMM similar to the original methods.
Fig.~\ref{fig:topology} shows an example of geodesic distance on non-zero genus surfaces computed using the face-based solver with the new formulation. The addition of global integrability conditions leads to more accurate geodesic distance, but the improvement is minor. In fact, in all our experiments, the original formulation already produces results that are close to the exact geodesic distance even for surfaces of complex topology, leaving little room of improvement. Although without a formal proof, we believe this is because the unit vector fields derived from heat diffusion are already close to the gradients of exact geodesic distance~\cite{CraneWW13}. Therefore, in practice the simple compatibility conditions~\eqref{eq::GradCompatibility} or~\eqref{eq:Integrability} are often enough to prevent pathological cases such as the cylinder example. Unless stated otherwise, all our results are generated using only constraints~\eqref{eq::GradCompatibility} or~\eqref{eq:Integrability}  to enforce integrability. 

   \begin{figure}[t]
   	\centering
   	\includegraphics[width=0.95\columnwidth]{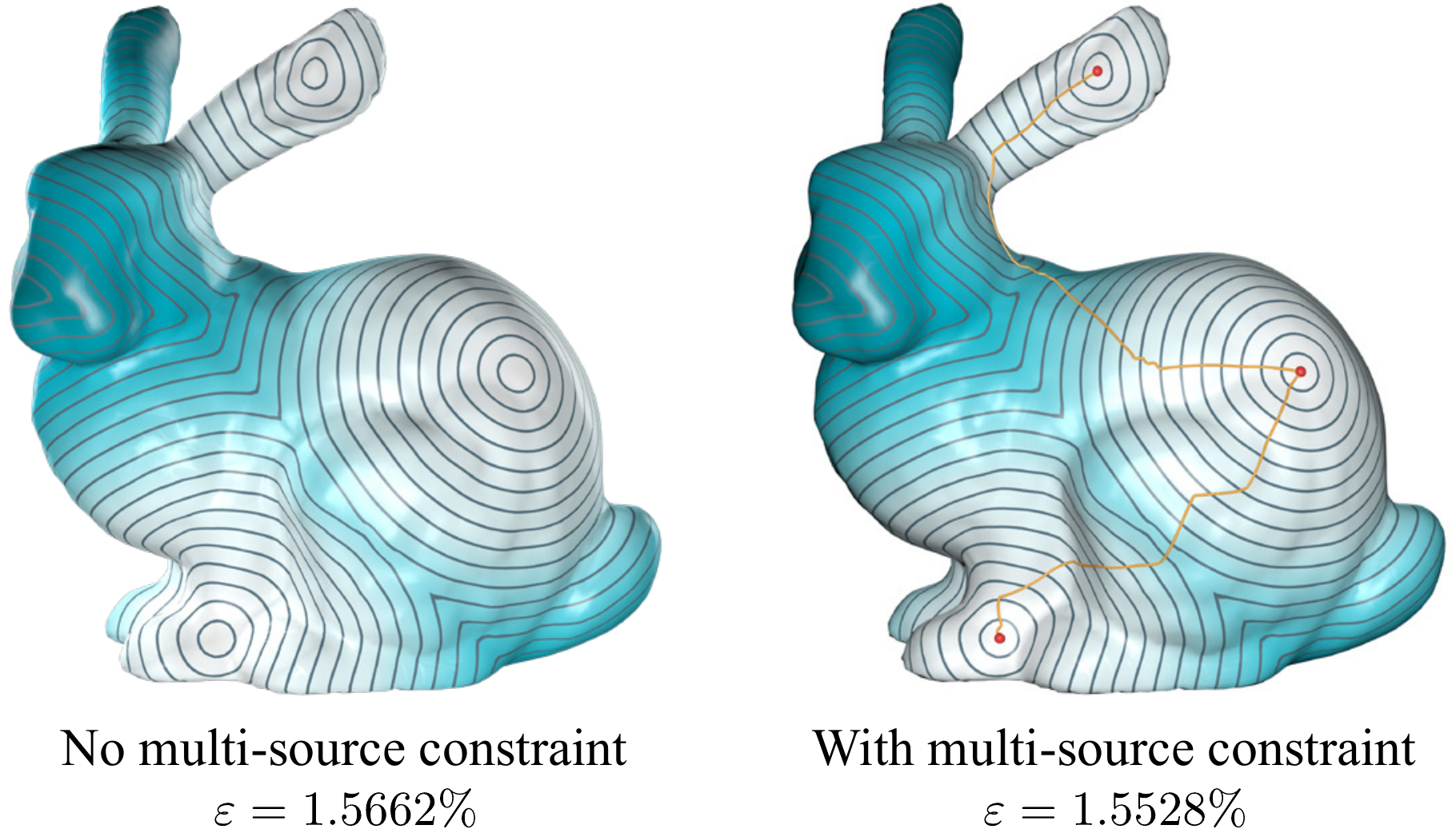}
   	\caption{
   		Geodesic distance from multiple sources, computed using our face-based method with and without the constraint~\eqref{eq:MultiSourceConstraint}, and its mean relative error $\varepsilon$ compared to the ground truth as defined in Eq.~\eqref{eq:MeanRelError}.
   	}
   	\label{fig:mSources}
   \end{figure}

\subsection{Multiple Sources}
Our method can be easily extended to the case of multiple sources. Following~\cite{CraneWW13}, we compute an initial unit vector field via heat diffusion from a generalized Dirac over the source set. This leads to a linear system with the same matrix as~\eqref{eq:heat} and a different right-hand-side. To adapt our single-source heat diffusion solver to this new linear system, we first construct the breadth-first vertex sets $\{\mathcal{D}_i\}$ by including all source vertices into $\mathcal{D}_0$ and collecting $\mathcal{D}_j$ ($j \geq 1$) according to the same definitions as in Eq.~\eqref{eq:VertexSets}; then the Gauss-Seidel iteration proceeds exactly the same as Section~\ref{sec:heat}. 
Afterward, we find the closest integrable gradient field $\{\mathbf{g}_i\}$ and integrate it to recover the geodesic distance, in the same way as Sections~\ref{sec:integrable} and \ref{sec:integration}. Strictly speaking, when computing $\{\mathbf{g}_i\}$ we need to enforce an additional constraint: since all source vertices have the same geodesic distance, the line integral of $\{\mathbf{g}_i\}$ along any path connecting two source vertices must vanish. To do so, we find the shortest path $\mathcal{S}$ along edges from one source to each of the other sources, and introduce one of the following constraints for $\mathcal{S}$:
\begin{itemize}
	\item For the face-based method, we require
	\begin{equation}
	\sum_{e_j \in \mathcal{S}} \mathbf{e}_j \cdot \mathbf{g}_{e_j} = 0,
	\label{eq:MultiSourceConstraint}
	\end{equation}
	where $e_j$ is an edge on the path, $\mathbf{e}_j$ is its vector in the same orientation as the path, and $\mathbf{g}_{e_j}$ is the gradient variable on a face adjacent to $e_j$.
	\item For the edge-based method, we enforce
		\begin{equation}
	\sum_{e_j \in \mathcal{S}} s_{e_j}^{\mathcal{S}} \cdot x_{e_j} = 0,
	\label{eq:MultiSourceConstraintEdgeBased}
	\end{equation}
	where $s_{e_j}^{\mathcal{S}} \in \{-1, 1\}$ indicates the orientation of edge $e_i$ with respect to the path.
\end{itemize}
The new optimization problem remains convex and is solved using ADMM. In our experiments, however, adding such constraints only makes a slight improvement to the accuracy of the results compared to the original formulation (see Fig.~\ref{fig:mSources} for an example with the face-based method). Again, this is likely because the unit vector field derived from heat diffusion is already close to the gradient of the exact geodesic distance. Moreover, our method for recovering geodesic distance always sets the distance at source vertices to zero, which ensures correct values at the sources even if the constraints~\eqref{eq:MultiSourceConstraint} or~\eqref{eq:MultiSourceConstraintEdgeBased} are violated. 

\begin{figure}[t]
	\centering
	\includegraphics[width=\columnwidth]{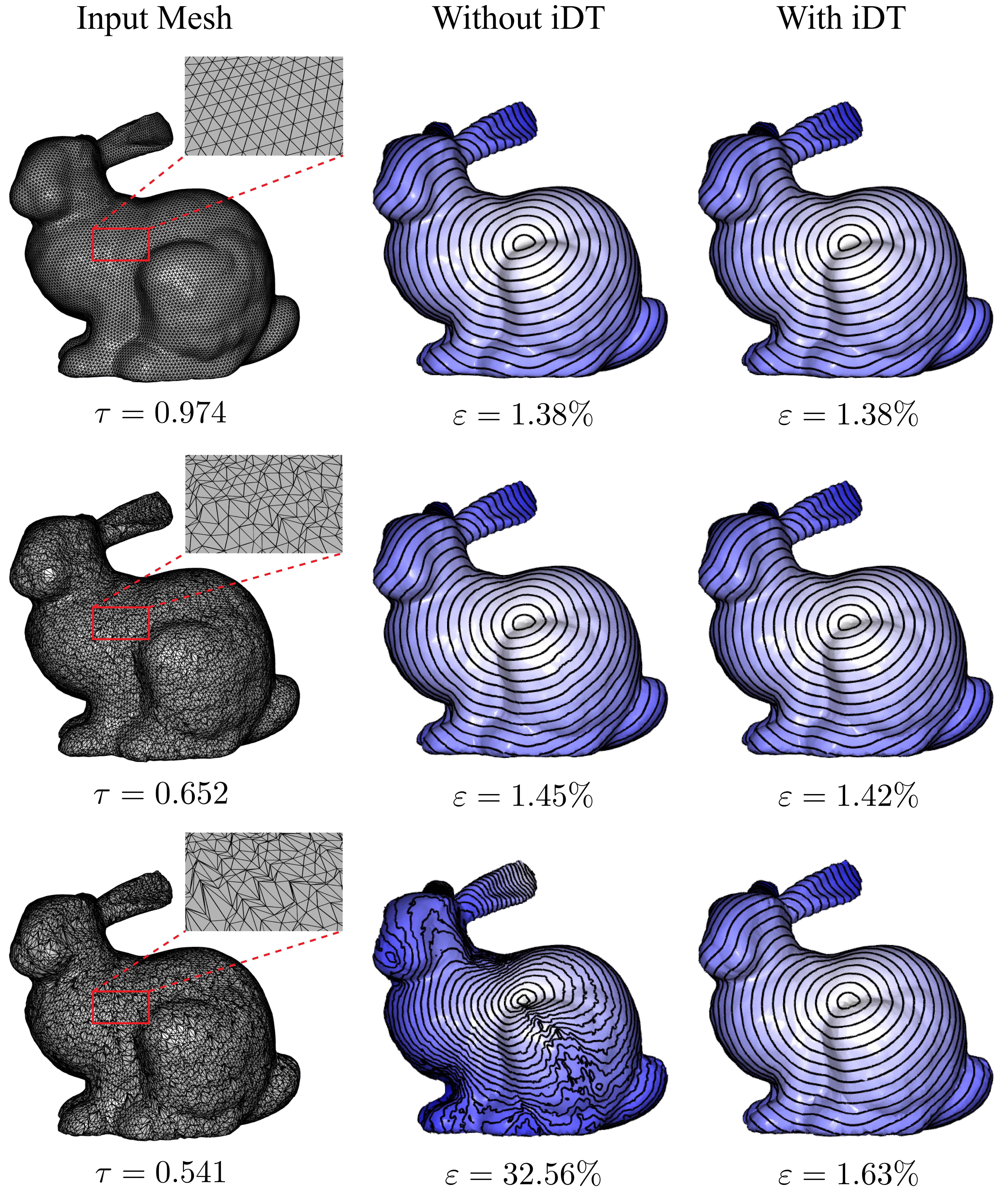}
	\caption{Given three input meshes with the same underlying geometry and different triangulation quality, using our edge-based method together with the intrinsic Delaunay triangulation helps to retain the accuracy of the computed geodesic distance even if the input mesh is poorly triangulated. Here the triangulation quality $\tau$ is defined in Eq.~\eqref{eq:TriangulationQuality}, and the mean relative error $\varepsilon$ compared to the ground truth as defined in Eq.~\eqref{eq:MeanRelError}.}
	\label{fig:IDT}
\end{figure} 

\begin{table*}[htbp]
	\caption{Comparison of computational time (in seconds), peak memory consumption (in MB), and accuracy between VTP~\cite{QH2016}, heat method~\cite{Crane2017}, variational heat method~\cite{belyaev2015}, and our methods (face-based and edge-based), on a PC with an octa-core CPU and 128GB memory. The accuracy is measured with the mean relative error $\varepsilon$ defined in Eq.~\eqref{eq:MeanRelError} using the VTP result as ground truth. For the heat method and its variants, we search for each model the optimal smoothing factor $m$ (see Eq.~\ref{eq:DiffusionTime}) that produces results with the best accuracy.}
	\label{tab:results}
	
	\renewcommand\arraystretch{1.4}
	\centering
	\setlength\tabcolsep{2pt}
	\begin{footnotesize}
		\resizebox{179mm}{47mm}{
			\begin{tabular}{!{\vrule width0.8pt}c|c||c|c|c|c|c|c|c|c|c|c|c!{\vrule width0.8pt}}
				\Xhline{0.8pt}
				\multicolumn{2}{!{\vrule width0.8pt}c||}{Model}              &Triomphe	&Connector	&ChineseLion	&Tricep	    &WelshDragon	&Lucy	        &HappyGargoyle	 &Cyvasse	    &TwoHeadedBunny	&MalteseFalcon	&ThunderCrab \\ \hline
				\multicolumn{2}{!{\vrule width0.8pt}c||}{Number of Vertices}                &997,635    &2,002,322	&3,979,442	    &7,744,320	&9,884,764	    &16,092,674 	&19,860,482	     &39,920,642	&40,103,938	       &79,899,650	    &98,527,234 \\ \Xhline{0.8pt}
				\multirow{2}{*}{VTP} &Time             &17.66	    &224.82	    &235.78	        &178.41	    &568.54	        &1320.15	&5060.72	     &12363.21	    &16681.46	          &28309.32	    &22298.76 \\ \cline{2-13}
				&RAM              &1,218	  &2,502  &4,849	 &9,664	  &12,394	&18,670	  &23,464	&46,976	  &47,258	&93,127	  &114,610 \\ \Xhline{0.8pt}
				\multirow{5}{*}{HM}  &Precompute        &10.22	  &26.17  &54.69	 &115.97  &153.34	&281.97	  &388.87	&\multicolumn{4}{c!{\vrule width0.8pt}}{Out of Memory} 	\\ \cline{2-13}
				& Solve &0.69	 &1.54	  &3.07  	&5.71	 &7.24	  &12.69	&16.46	 &\multicolumn{4}{c!{\vrule width0.8pt}}{Out of Memory}	\\ \cline{2-13}
				&RAM              &2,951	  &6,160  &12,207   &23,945	  &30,677	&51,144   &65,545   &\multicolumn{4}{c!{\vrule width0.8pt}}{Out of Memory}	\\ \cline{2-13}
				&$\varepsilon$    &0.31\%  &0.15\%	&0.73\%	 &0.84\%  &1.21\%	&0.46\%	  &0.29\%	 &\multicolumn{4}{c!{\vrule width0.8pt}}{Out of Memory}	\\	 \cline{2-13}
				&m                &1.0	      &2.3	    &3.1	 &12.6	  &11.1	    &13.0	      &15.7	    &\multicolumn{4}{c!{\vrule width0.8pt}}{Out of Memory}	\\ \Xhline{0.8pt}
				\multirow{5}{*}{Variational-HM} & Precompute &11.7	 &29.34	&61.31	 &129.07  &171.24	&310.01	  &424.81	&\multicolumn{4}{c!{\vrule width0.8pt}}{Out of Memory}	\\	\cline{2-13}
				&Solve            &5.55	  &11.69	&23.78	 &48.61 	&63.67	&113.99	  &149.45    &\multicolumn{4}{c!{\vrule width0.8pt}}{Out of Memory}    \\ \cline{2-13}
				&RAM              &3,265   &6,805	&13,426	 &26,377	&33,720	&56,186	  &71,644	&\multicolumn{4}{c!{\vrule width0.8pt}}{Out of Memory}	 \\ \cline{2-13}
				&$\varepsilon$    &0.44\%  &0.14\%	&0.52\%	 &0.59\%	&0.91\%	&0.29\%	  &0.21\%	&\multicolumn{4}{c!{\vrule width0.8pt}}{Out of Memory}	 \\ \cline{2-13}
				&m                &1.0	      &2.3  	&3.4     &13.4	    &11.8	&13.9     &16.6	    &\multicolumn{4}{c!{\vrule width0.8pt}}{Out of Memory}	\\ \Xhline{0.8pt}
				\multirow{4}{*}{Ours (Face)} &Time            & 3.44   	&11.29	 &20.29	    &50.24	 &65.20	    &148.65	    &171.73	    &355.42	  &343.07	&834.52	  &1174.45 \\ \cline{2-13}
				&RAM             &1,071  	&1,974	 &3,828	    &7,417	 &9,420	    &15,370	    &18,957  	&38,027	  &38,202	&76,106	  &93,850 \\ \cline{2-13}
				&$\varepsilon$   &0.48\%	    &0.59\%	 &0.46\%	&0.50\%	 &0.50\%	&0.50\%	    &0.55\%  	&0.86\%	  &0.83\%	&0.82\%	  &1.24\% \\ \cline{2-13}
				&GS Iters        &300	    &800     &800 	    &1000	 &1200	    &1500        &2000	    &2000	  &2000	    &2000	  &2000 \\ \cline{2-13}
				&m               &1.0	    &1.0	 &1.0	    &1.0     &1.0	    &1.0	    &1.0	    &1.0      &1.0   	&1.0      &1.0 \\ \Xhline{0.8pt}
				\multirow{4}{*}{Ours (Edge)} &Time            &2.06	    &5.60	 &10.13  	&33.37	 &37.32	    &80.15	    &87.86	    &311.33	  &298.25	&1283.56  &1657.98 \\ \cline{2-13}
				&RAM             &543	    &983	 &1,942     &3,777	 &4,819	    &7,872	    &9,704	    &19,476   &19,566	&38,933	  &48,009 \\ \cline{2-13}
				&$\varepsilon$   &0.77\%	    &0.81\%	 &0.96\%	&0.84\%	 &1.00\%	&0.86\%	    &1.05\%	    &0.93\%	  &0.94\%	&0.94\%	  &1.41\% \\ \cline{2-13}
				&GS Iters        &200	    &400	 &400	    &700	 &700	    &800	    &1000	    &1900	  &1800	    &4000	  &4000 \\  \cline{2-13}
				&m               &1.0	    &1.0	 &1.0	    &1.0     &1.0	    &1.0	    &1.0	    &1.0      &1.0   	&10.0      &10.0 \\ \Xhline{0.8pt}
		\end{tabular}}%
	\end{footnotesize}
	
\end{table*}

\subsection{Intrinsic Delaunay Triangulation}
Like the original heat method, our methods are less accurate on meshes with triangulations of lower quality. The robustness and accuracy of heat methods can be improved by building the linear systems with respect to the \emph{intrinsic Delaunay triangulation} (iDT)~\cite{Bobenko2007} of the given mesh~\cite{Sharp2019}. Our methods also benefit from using iDT instead of the original mesh triangulation. The idea of intrinsic triangulation is that an edge connecting two vertices is a straight path on along the exact mesh surface instead of the ambient Euclidean space~\cite{Sharp2019-NIT}. 
Such intrinsic triangulation is represented using the vertex connectivity and the length of each edge, rather than the 3D coordinates of the vertices (i.e., their extrinsic embedding in the ambient space). 
Using such edge lengths, the intrinsic angles within each triangle can be easily computed by isometrically unfolding the triangle into the plane (i.e., preserving its edge lengths) and applying standard Euclidean formulas~\cite{Sharp2019-NIT}. 
An iDT is an intrinsic triangulation where each pair of neighboring triangles satisfy the intrinsic Delaunay property about their angles~\cite{Bobenko2007,Sharp2019-NIT}. It can be computed using the edge-flipping algorithm proposed in~\cite{Bobenko2007}. Despite its high worst-case complexity in theory, the edge-flipping algorithm is surprisingly efficient in practice and its computational cost often grows approximately linearly with respect to the mesh size~\cite{Sharp2019-NIT}.

Using iDT, the heat diffusion step still amounts to solving the linear system~\eqref{eq:heat}, with the vertex areas and the cotangent values in the system matrix computed from the intrinsic edge lengths. Our Gauss-Seidel heat diffusion can be easily applied on the iDT of a mesh, with the breadth-first vertex sets set up using the vertex connectivity from the iDT, and with the update formula~\eqref{eq:HeatDiffusionUpdate} evaluated using the intrinsic vertex areas and cotangent values. Using the heat diffusion solution $\mathbf{u}$, we isometrically unfold each triangle $f$ to evaluate the gradient of $\nabla u |_{f}$ on the triangle, and represent $\nabla u |_{f}$ using 2D coordinates with respect to a local frame defined on the unfolded triangle. From this local representation we apply the normalization in Eq.~\eqref{eq:HeatDiffusionNormalization} to derive the target gradients for the geodesic distance.

Our edge-based optimization approach can be easily extended to the iDT setting. From the target gradient on each triangle, we isometrically unfold the triangle to evaluate its target edge differences according to Eq.~\eqref{eq:TargetDifference}. Then the solver proceeds exactly the same as in Section~\eqref{sec:edgeformulation}. 

To apply our face-based method, we need to represent the gradient variables and the auxiliary variables using local 2D coordinates within each triangle, and rewrite the integrability constraint in~\eqref{eq::GradCompatibility} as a linear condition that involves transformation between the local frames on the two adjacent triangles. A similar ADMM solver can be derived for this intrinsic formulation.

Fig.~\ref{fig:IDT} compares the results using our edge-based method with and without iDT. The triangulation quality of the original mesh $\mathcal{M}$ is evaluated using the following measure~\cite{Pebay2003}:
\begin{equation}
	\tau(\mathcal{M}) = \frac{1}{|\mathcal{F}|} \sum_{f \in \mathcal{F}} \frac{2 \sqrt{3} \cdot R_f}{l_f},
	\label{eq:TriangulationQuality}
\end{equation}
where $R_f$ and $l_f$ are the inradius and the maximum edge length of a triangle $f$. A larger value of $\tau$ indicates better quality of the triangulation. We can see that as the triangulation quality worsens, iDT helps to retain the accuracy of the computed geodesic distance. In this paper, we do not employ iDT other than in Fig.~\ref{fig:IDT}.
\section{Experimental Results \& Discussions}\label{sec:results}

\begin{figure*}[t!]
	\centering
	\includegraphics[width=\textwidth]{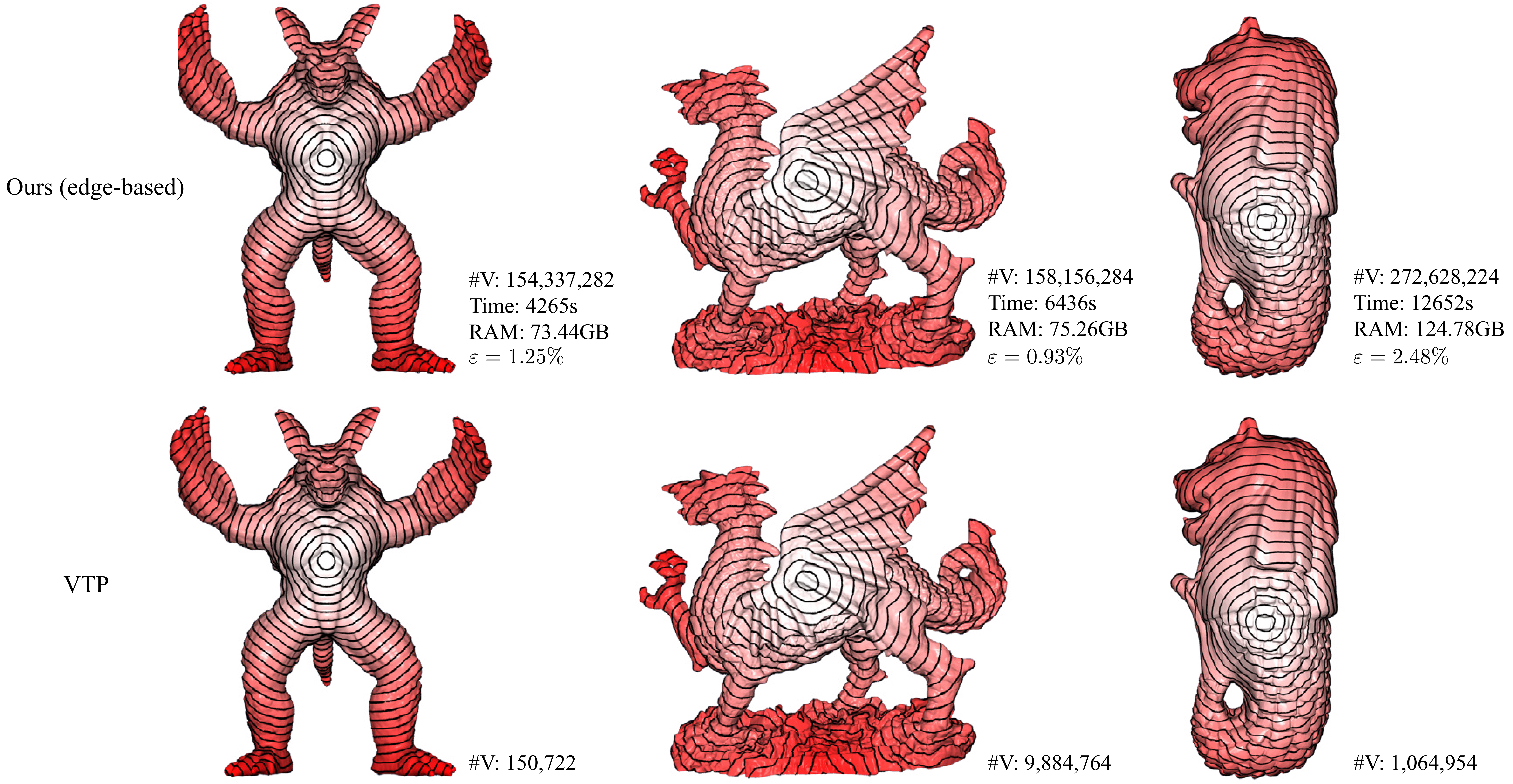}
	\caption{We repeatedly subdivide some meshes (bottom row) to obtain very high-resolution models (top row). On such models, even our face-based method runs out of memory, while our edge-based method can still work on a PC with 128GB RAM and an octa-core CPU at 3.6 GHz. The numbers on the top row show the performance and accuracy of our edge-based method on each model. The bottom row shows the VTP results on the original models for comparison.}
	\label{fig:edgebased}
\end{figure*}

We implement our algorithm in C++ and use OpenMP for parallelization. The source code is available at \url{https://github.com/bldeng/ParaHeat}.
Following~\cite{Crane2017}, we set the heat diffusion time as
\begin{equation}
t = m \cdot h^2,
\label{eq:DiffusionTime}
\end{equation}
where $h$ is the average edge length and $m$ is a smoothing factor. In our experiments, setting $m$ to a value between $1$ and $10$ leads to good results.
For the maximal iteration of the ADMM solver $I_{\max{}}$, we observe a good balance between speed and accuracy by setting $I_{\max{}} = 10$, and use it as the setting in all our experiments.

\subsection{Performance Comparison} 

In the following, we evaluate the performance of our algorithm (including the face-based method and edge-based method), and compare it with the state-of-the-art methods, including the original heat method (HM) with Cholesky decomposition~\cite{Crane2017}, the variational heat method (Variational-HM)~\cite{belyaev2015}, and the VTP method~\cite{SVG}. 
Tab.~\ref{tab:results} reports the running time, peak memory consumption and accuracy of each method, for the models in Fig.~\ref{fig:gallery}.
All examples are run on a desktop PC with an Intel Core i7 octa-core CPU at 3.6GHz and 128GB of RAM.
For the original heat method, we show its pre-computation time for matrix factorization and the solving time for back substitution separately. For our method, we show the computational time using eight threads. 
For a fair comparison, for both HM and Variational-HM the linear system matrix factorization is done using the MA87 routine from the HSL library~\cite{HSL}, which is a multi-core sparse Cholesky factorization method~\cite{Hogg10}. Following~\cite{Surazhsky05}, we measure the accuracy of each method using its mean relative error $\varepsilon$ defined as
\begin{equation}
	\varepsilon = \frac{1}{|\mathcal{V}|} \sum_{v \in \mathcal{V} \setminus \mathcal{S}} \frac{|d(v) - d^\ast(v)|}{|d^\ast(v)|},
	\label{eq:MeanRelError}
\end{equation}
where $\mathcal{V}$ is the set of mesh vertices, $\mathcal{S}$ is the set of source vertices, and $d^\ast(v), d(v)$ are the ground truth geodesic distance and the distance computed by the method at vertex $v$, respectively. Since VTP computes the exact geodesic distance, we use its result as the ground truth.
From Tab.~\ref{tab:results} we can see that our methods consume less memory than all other methods, and take the least computational time while achieving similar accuracy as the heat method. The difference is the most notable on the largest models with more than 39 million vertices: both the heat method and the variational heat method run out of memory, while our methods are more than an order of magnitude faster than VTP and produce results with mean relative errors no more than 1.41\%. Between our two approaches, the edge-based method consumes about 50\% less memory than the face-based method.

\begin{figure*}[t!]
	\centering
	\includegraphics[width=\textwidth]{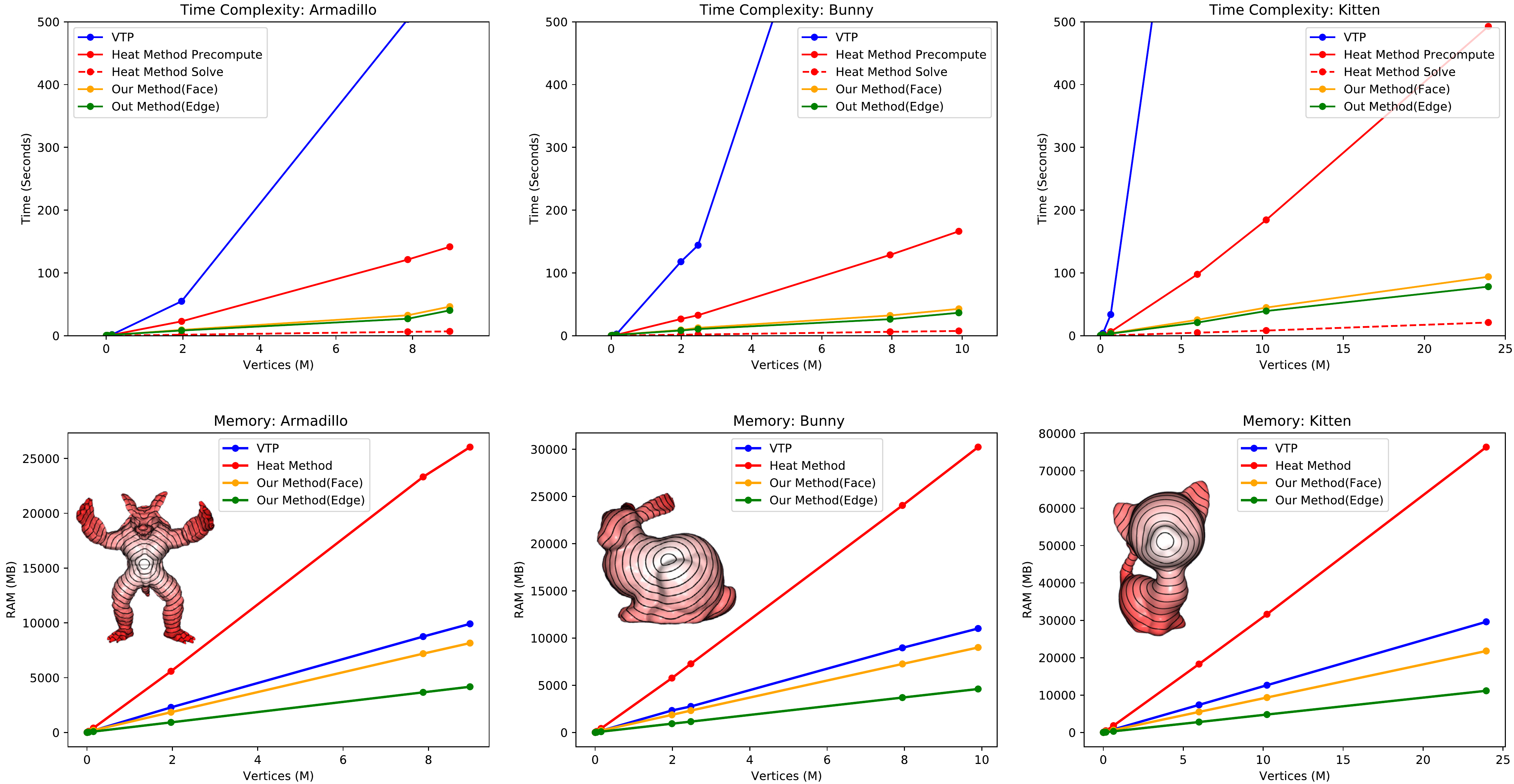}\\
	\caption{Time and space complexity. We show how the computational time (top row) and peak memory consumption (bottom row) change along with different mesh resolutions on three test models, using our methods, VTP~\cite{QH2016}, the heat method~\cite{Crane2017} and the variational heat method~\cite{belyaev2015}, respectively.}
	\label{fig:complexities}
\end{figure*}

\begin{wrapfigure}{r}{0.3\columnwidth}
	\vspace*{-6pt}
	\centering
	\includegraphics[width=0.3\columnwidth]{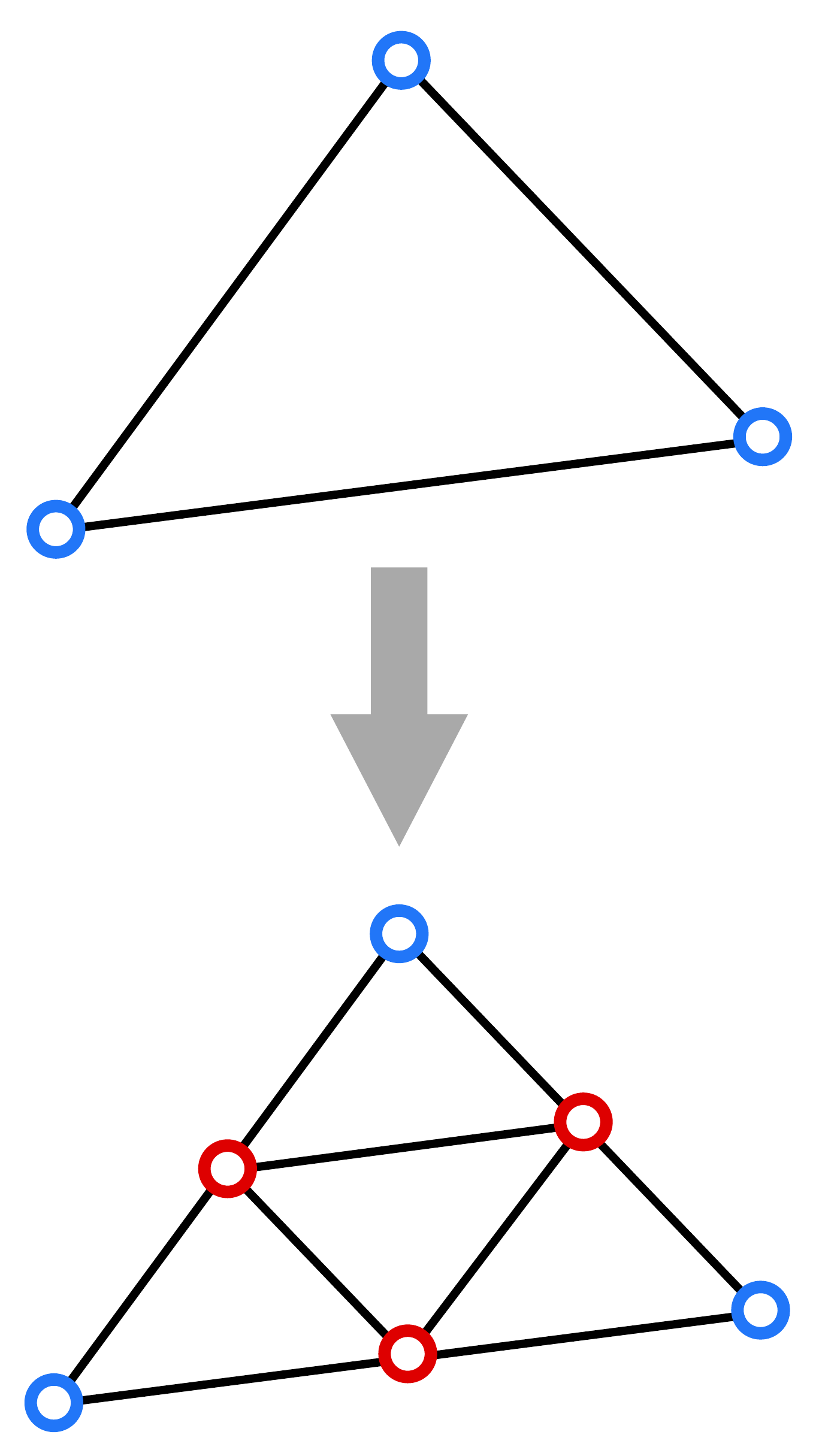}
	\vspace*{-12pt}
\end{wrapfigure}
To further verify the memory efficiency of our edge-based method, we apply it to models with even higher resolution than those in Tab.~\ref{tab:results}. One challenge for such tests is to measure the accuracy, because VTP will also run out of memory on such models and cannot provide the ground truth. To overcome this problem, we take some high-resolution meshes where VTP is applicable, and repeatedly subdivide them to increase their resolution. 
Each subdivision step splits each triangle into four triangles, by adding a new vertex at the midpoint of each edge and then connecting these new vertices, as shown in the inset. Note that such operation does not alter the metric on the mesh surface, as the four new triangles coincide with the original triangle. Therefore, if we take a vertex $v_s$ that exists in the initial mesh $\mathcal{M}^0$ as the source vertex, then for any other vertex $v$ that also exists in $\mathcal{M}^0$, their geodesic distance on the initial mesh and on a subdivided mesh has the same value. Therefore, to measure the accuracy, we select a vertex from $\mathcal{M}^0$ as the source to compute the geodesic distance on the subdivided mesh, and evaluate its mean relative error using only the vertices that exist in the initial mesh, with the VTP solution on $\mathcal{M}^0$ as the ground truth. Fig.~\ref{fig:edgebased} shows some examples of such models, where even our face-based method runs out of memory, while our edge-based method can produce results with good accuracy.

\subsection{Space and Time Complexity} 
Fig.~\ref{fig:complexities} compares the growth of computational time and peak memory consumption between VTP, the original heat method, and our methods, on meshes with the same underlying geometry in different resolutions. The graphs verify the $O(n^2)$ time complexity of VTP as well as the fast growth of memory footprint for the original heat method, which cause scalability issues for both methods. In comparison, our methods achieve nearly linear growth in both computational time and peak memory consumption, allowing them to handle much larger meshes.

\subsection{Parallelization} 
\begin{figure}[!t]
	\centering
	\includegraphics[width=\columnwidth]{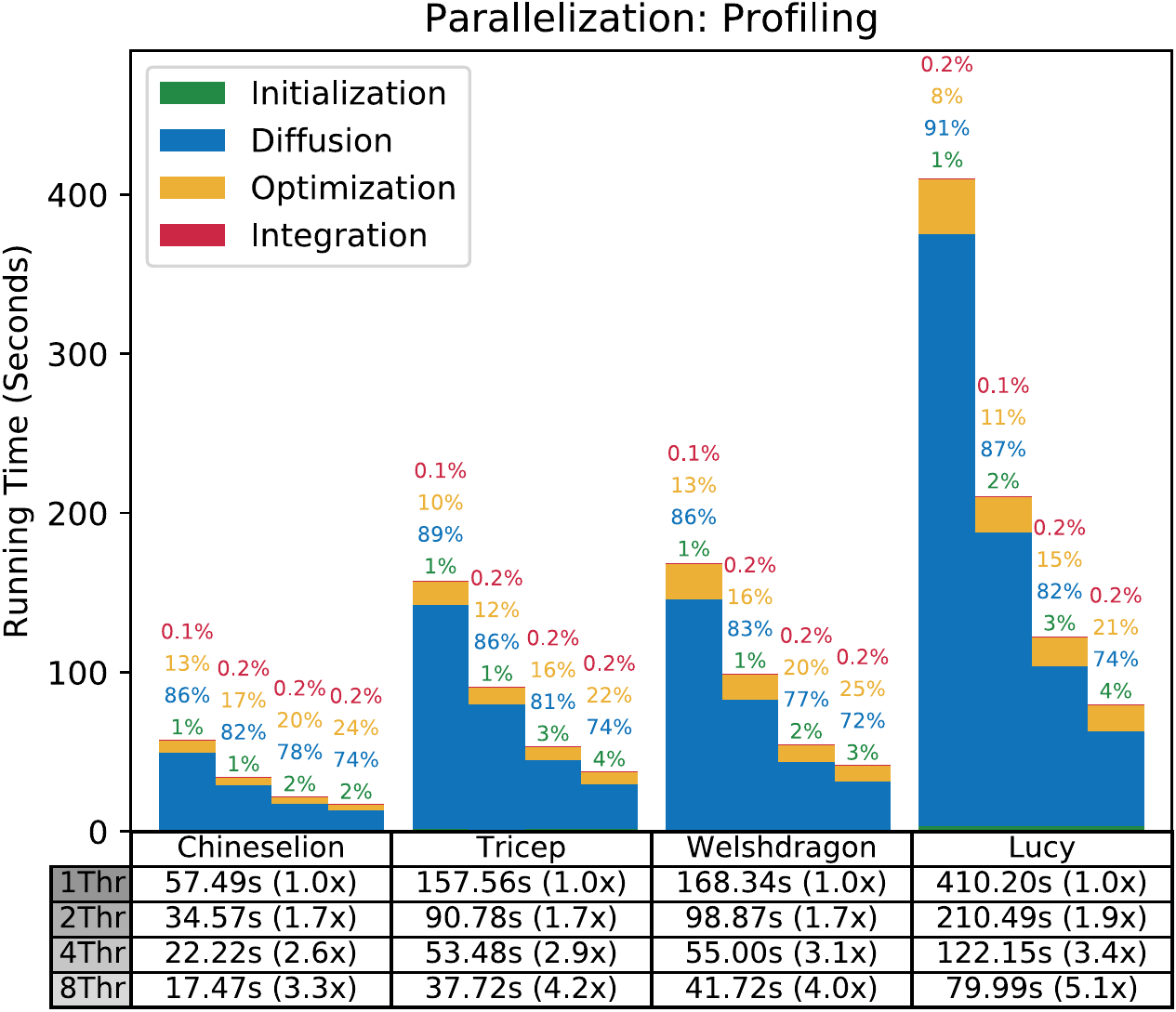}
	\caption{Timings for the four steps of our face-based method: initialization of data structures, heat diffusion, gradient optimization, and gradient integration. The numbers on top of each column show the percentage of timing spent on each step. The timings are measured on four models using one, two, four, and eight threads, respectively. All examples are tested with 600 Gauss-Seidel iterations and 10 ADMM iterations.}
	\label{fig:parallel}
\end{figure}

To evaluate the speedup from parallelization, we re-run our face-based method on the same model using one, two, four, and eight threads, respectively. We fix the Gauss-Seidel iteration count to 600 in the experiments. Fig.~\ref{fig:parallel} shows the timing for each configuration and the percentage spent on each of the following four steps: initialization of data structures, heat diffusion, gradient optimization, and gradient integration. We observe that the total timing is approximately inversely proportional to the number of threads, indicating a high level of parallelism and low overhead of our method. Although some initialization steps such as breadth-first traversal of vertices are performed sequentially, they only contribute a small percentage of the total timing and incur no bottleneck to the overall performance. As our edge-based method follows the same algorithmic principle, its parallelization performance is similar to the face-based method.

\subsection{Comparison with Iterative Linear Solver} 
\begin{figure}[t!]
	\centering
	\includegraphics[width=1\columnwidth]{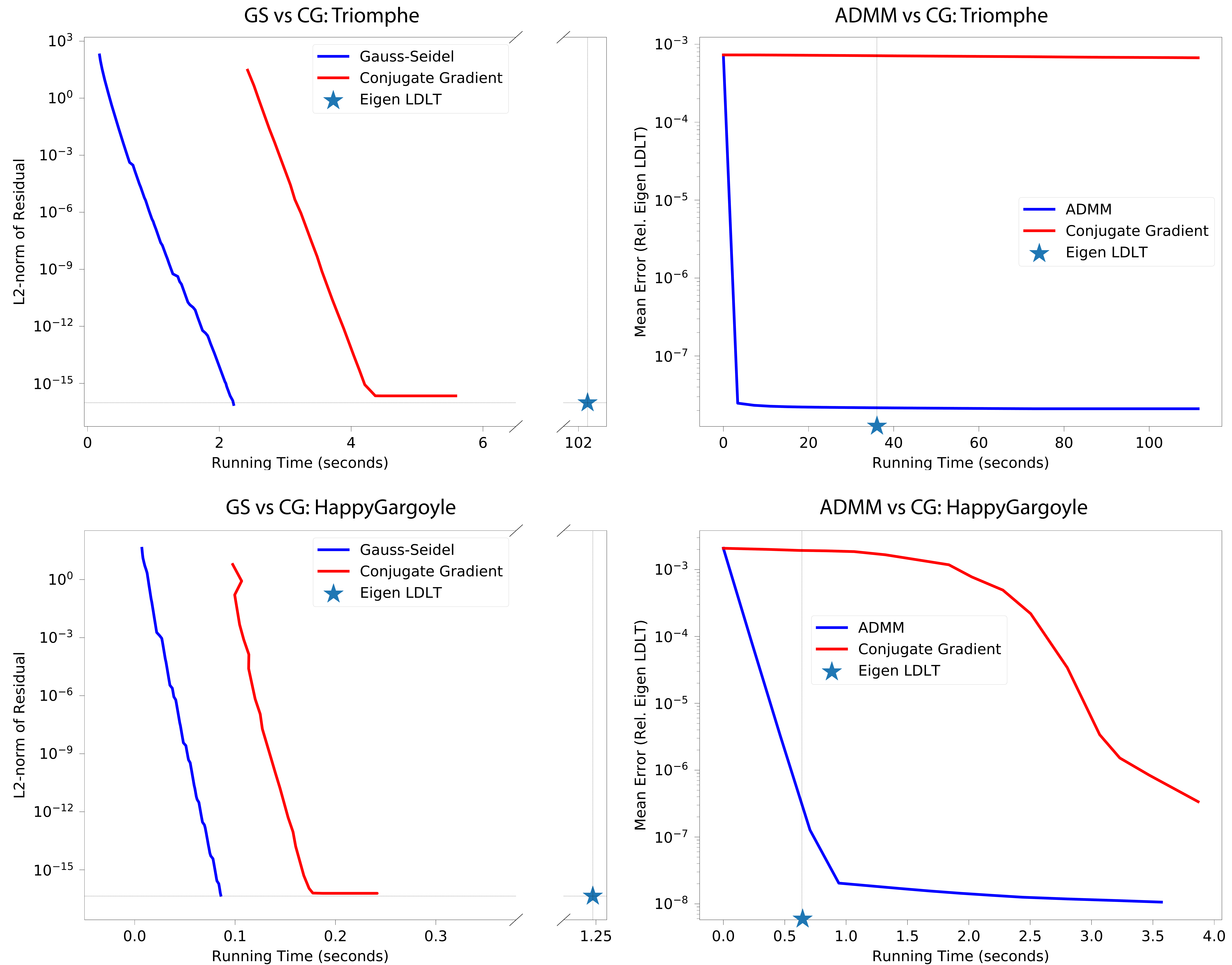}
	\caption{Comparison between our face-based method and the original heat method using a conjugate gradient solver. The graphs show the change of solution accuracy with respect to computational time. The solution accuracy for the heat diffusion and the geodesic distance recovery is computed according to Eqs.~\eqref{eq:HeatSolutionError} and \eqref{eq:GeodDistSolutionError}, respectively.}
	\label{fig:cg}
\end{figure}

The original heat method does not scale well because it adopts a direct linear solver that is known to have high memory footprints for large-scale problems. It is worth noting that the sparse linear systems in the original heat method can also be handled using iterative linear solvers with better memory efficiency for large problems. To evaluate the effectiveness of such approaches compared with our methods, we replace the direct linear solver in the original heat method with the conjugate gradient (CG) method, a popular iterative solver for sparse positive definite linear systems. As the Poisson system in the heat method is only positive semi-definite, we fix the solution components at the source vertices to zero and reduce the system to a positive definite one that can be solved using CG. We adopt the CG solver from the \textsc{Eigen} library~\cite{eigenweb} using the diagonal preconditioner. In Fig.~\ref{fig:cg}, we show the relation between accuracy and computational time using the CG solver for heat diffusion and geodesic distance recovery, and compare it with our Gauss-Seidel (GS) heat diffusion and face-based ADMM gradient solver. For heat diffusion, we measure the accuracy of a solution $\mathbf{u}$ using the $\ell_2$-norm of its residual with respect to the heat diffusion equation~\eqref{eq:heat}:
\begin{equation}
	E_1(\mathbf{u}) = \|(\mathbf{A}-t \mathbf{L}_c) \mathbf{u} -\mathbf{u}_0 \|_2.
	\label{eq:HeatSolutionError}
\end{equation}
For geodesic distance recovery, we measure the accuracy of a solution $\mathbf{d}$ using the $\ell_2$-norm of its difference with the solution $\mathbf{d}^\ast$ from a direct solver:
\begin{equation}
	E_2(\mathbf{d}) = \frac{1}{n}\|\mathbf{d} - \mathbf{d}^\ast\|_2,
	\label{eq:GeodDistSolutionError}
\end{equation}
where $n$ is the number of components in $\mathbf{d}$. We compute $\mathbf{d}^\ast$ using the LDLT solver from \textsc{Eigen}. For the ADMM solver, we integrate the resulting gradients according to Section~\ref{sec:integration} to obtain the solution $\mathbf{d}$. For reference, we also include the computational time and accuracy of the \textsc{Eigen} LDLT solver in the comparison. Fig.~\ref{fig:cg} shows that both our GS heat diffusion and our ADMM gradient solver improves the accuracy faster than CG, making our approach a more suitable choice for handling large models.

\section{Discussion \& Future Work}
\label{sec:conclusion}

In this paper, we develop a scalable approach to compute geodesic distance on mesh surfaces. We adopt a similar approach as the heat method, first approximating the geodesic distance gradients via heat diffusion, and then recovering the distance by integrating a corrected gradient field. Unlike the heat method that directly solves two linear systems, we propose novel algorithms that can be easily parallelized and with fast convergence. Our approach significantly outperforms the heat method while producing results with comparable accuracy. Moreover, its memory consumption grows linearly with respect to the mesh size, allowing it to handle much larger models. We perform extensive experiments to evaluate its speed, accuracy, and robustness. The results verify the efficiency and scalability of our methods. 

Our method can be improved and extended in several aspects. First, although ADMM converges quickly to a solution of moderate accuracy, it can take a large number of iterations to converge to a highly accurate solution~\cite{Boyd2011-admm}. One possible way to tackle this issue is to employ the accelerated ADMM solver proposed in~\cite{Goldstein2014}, which often converges faster but at the cost of higher memory consumption and extra computation per iteration for checking the decrease of residuals. 
Another potential approach is to switch to another solver with faster local convergence when the ADMM convergence slows down. One such candidate is L-BFGS, which maintains the history of $m$ previous iterations so that its memory consumption also grows linearly with respect to the mesh size. On the other hand, each L-BFGS iteration involves line search that is potentially time-consuming, and fast convergence may require a large value of $m$ that can still incur high memory footprint. A more in-depth investigation of these approaches will be performed in the future.

The convergence speed of our ADMM solvers is also affected by the penalty parameter, as well as the scaling of linear side constraints that serves as pre-conditioning (e.g., the matrix $\mathbf{M}$ in constraint~\eqref{eq:ADMMMatrixFormConstraint}). Although our current choice of such parameters works well, it is possible to find better parameter values that can further improve the convergence. However, existing methods for selecting such optimal parameters either only work for simple problems~\cite{Ghadimi2015}, or involve time-consuming computation that potentially negates the benefit of faster convergence~\cite{Giselsson2017}. A practical approach for choosing optimal parameters is worth further investigation.

For the original heat method, we have shown that solving the linear systems with conjugate gradient suffers from slow convergence. This issue can potentially be resolved using multigrid methods. However, such methods bring additional costs in computational time and memory consumption for building the multigrid hierarchy, as well as the question of how to build a good hierarchy on unstructured meshes. Finding a suitable multigrid method is an interesting topic for future work. Also worth investigating is how our ADMM solver can be adapted to benefit from multigrid hierarchies.

Despite experiments showing that our formulation can handle surfaces of non-zero topology and multiple sources using only local compatibility of the gradients, we do not have a formal proof of this property. An interesting avenue for future research is how the accuracy of heat flow gradients affects the effectiveness of our formulation. 

Our approach is currently limited to triangle meshes because of its reliance on a well-defined discrete gradient operator. As an extension, we would like to explore its application on other geometric representations such as point clouds and implicit surfaces.  

Finally, Although we only implement our method on CPUs using OpenMP, its massive parallelism allows it to be easily ported to GPUs, which we will leave as future work.

\section*{Acknowledgments}
The authors are supported by National Natural Science Foundation of China (No. 61672481), Youth Innovation Promotion Association CAS (No. 2018495), and Singapore MOE RG26/17.

\ifCLASSOPTIONcaptionsoff
  \newpage
\fi



%
\bibliographystyle{IEEEtran}
\bibliography{Geodesic}

%

\vfill

\begin{IEEEbiography}[{\includegraphics[width=1in]{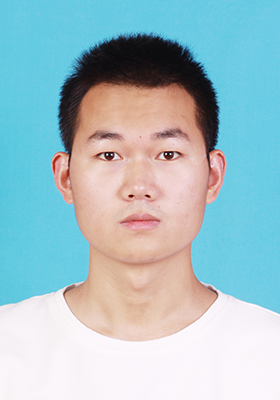}}]{Jiong Tao} obtained his master degree in Mathematical Sciences from University of Science and Technology of China in 2019, and the bachelor degree in School of the Gifted Young from same university in 2016. His research interests include computer graphics, geometry processing and numerical optimization.
\end{IEEEbiography}

\begin{IEEEbiography}[{\includegraphics[width=1in]{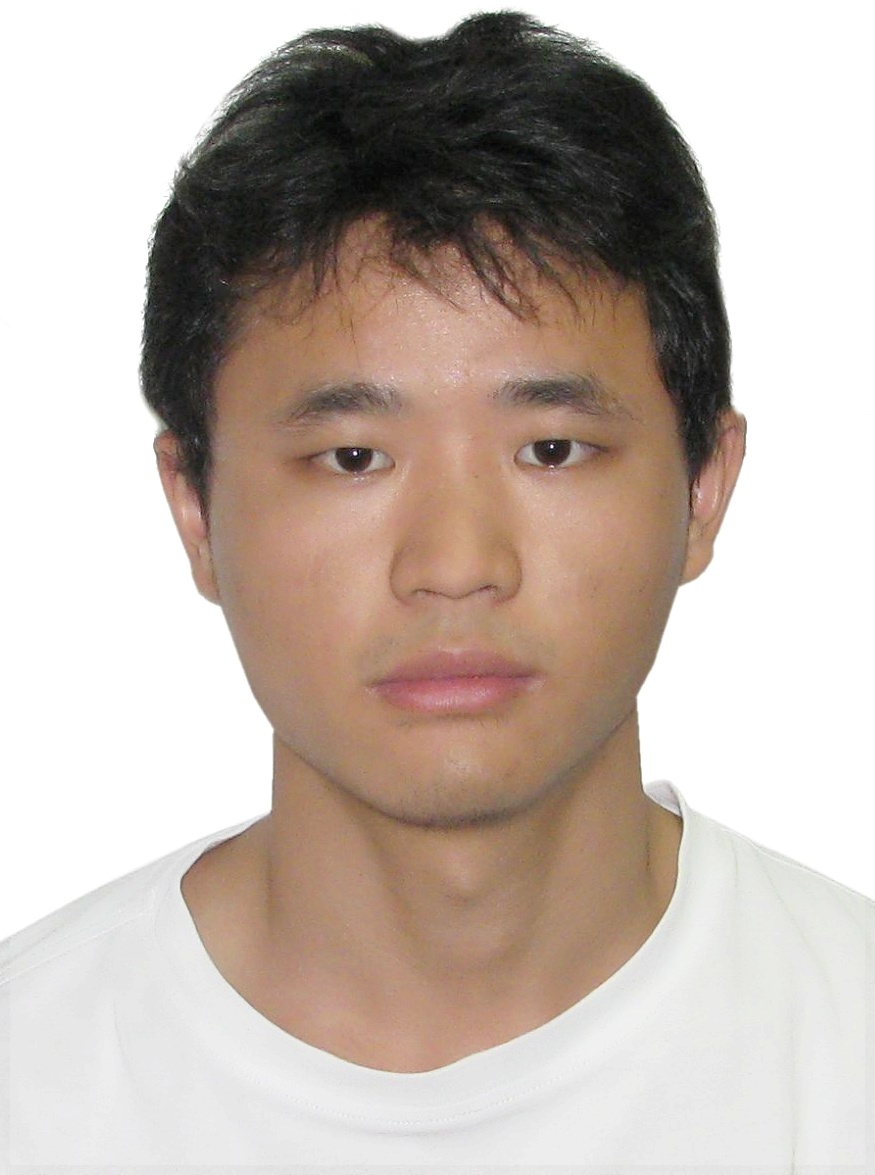}}]{Juyong Zhang}
is an associate professor in the School of Mathematical Sciences at University of Science and Technology of China. He received the BS degree from the University of Science and Technology of China in 2006, and the PhD degree from Nanyang Technological University, Singapore. His research interests include computer graphics, computer vision, and numerical optimization. He is an associate editor of The Visual Computer.
\end{IEEEbiography}

\begin{IEEEbiography}[{\includegraphics[width=1in]{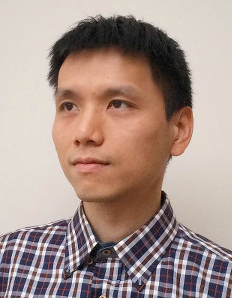}}]{Bailin Deng}
is a lecturer in the School of Computer Science and Informatics at Cardiff University. He received the BEng degree in computer software (2005) and the MSc degree in computer science (2008) from Tsinghua University (China), and the PhD degree in technical mathematics from Vienna University of Technology (Austria). His research interests include geometry processing, numerical optimization, computational design, and digital fabrication. He is a member of the IEEE.
\end{IEEEbiography}

\begin{IEEEbiography}[{\includegraphics[width=1in]{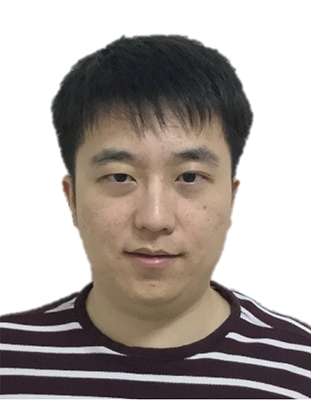}}]{Zheng Fang} received the BS degree in computer science and technology from Tianjin University, China. He is currently working toward the PhD degree with the School of Computer Science and Engineering, Nanyang Technological University, Singapore. His current research interests include computational geometry and computer graphics.
\end{IEEEbiography}

\begin{IEEEbiography}[{\includegraphics[width=1in]{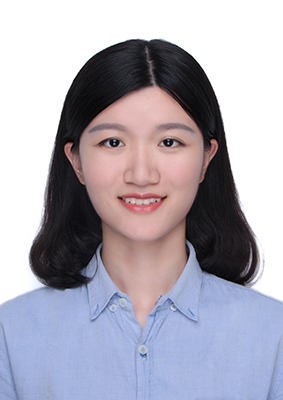}}]{Yue Peng} received the BS degree from the University of Science and Technology of China in 2016. Currently, she is working toward a PhD degree in Mathematical Sciences from University of Science and Technology of China. Her research interests include computer graphics and geometry processing.
\end{IEEEbiography}

\begin{IEEEbiography}[{\includegraphics[width=1in]{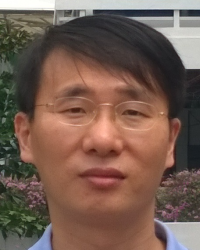}}]{Ying He} is an associate professor at School of Computer Science and Engineering, Nanyang Technological University, Singapore. He received the BS and MS degrees in electrical engineering from Tsinghua University, China, and the PhD degree in computer science from Stony Brook University, USA. His research interests fall into the general areas of visual computing and he is particularly interested in the problems which require geometric analysis and computation. For more information, visit http://www.ntu.edu.sg/home/yhe/
\end{IEEEbiography}


\vfill


\end{document}